%% file: 000-paper.tex
\documentclass[manuscript]{acmart}

\usepackage{soul}
\usepackage{enumitem}
\usepackage{rotating}
\usepackage{multirow}
\usepackage{xcolor, colortbl}
\usepackage{tcolorbox}
\usepackage{diagbox}
\usepackage{wrapfig}
\usepackage{siunitx}
\usepackage{graphicx}
\usepackage{caption}
\usepackage{subcaption}
\usepackage{textcomp}
\usepackage{mathtools}
\usepackage{adjustbox}
\usepackage{listings}

\newcommand*{\noteON}{} 

\AtBeginDocument{%
  \providecommand\BibTeX{{%
    \normalfont B\kern-0.5em{\scshape i\kern-0.25em b}\kern-0.8em\TeX}}}

\setcopyright{none}

\begin{document}
\title{%
An LLM's Attempts to Adapt to Diverse Software Engineers' Problem-Solving Styles: More Inclusive \& Equitable?
}

\include{z-commands}

\include{z-stat-commands}
\author{Andrew Anderson}
\email{Andrew.Anderson2@ibm.com}
\orcid{0000-0003-4964-6059}
\affiliation{\institution{IBM Research}}

\author{David Piorkowski}
\email{djp@ibm.com}
\orcid{0000-0002-6740-4902}
\affiliation{\institution{IBM Research}}

\author{Margaret Burnett}
\email{burnett@eecs.oregonstate.edu}
\orcid{0000-0001-6536-7629}
\affiliation{\institution{Oregon State University}}

\author{Justin Weisz}
\email{jweisz@us.ibm.com}
\orcid{0000-0003-2228-2398}
\affiliation{\institution{IBM Research}}

\renewcommand{\shortauthors}{Anderson et al.}
\renewcommand{\shorttitle}{An LLM's Attempts to Adapt to Engineers' Problem-Solving Styles}

\begin{abstract}
Software engineers use code-fluent large language models (LLMs) to help explain unfamiliar code, yet LLM explanations are not adapted to engineers' diverse problem-solving needs. 
We prompted an LLM to adapt to five problem-solving style types from an inclusive design method, the Gender Inclusiveness Magnifier (GenderMag).
We ran a user study with software engineers to examine the impact of explanation adaptations on software engineers' perceptions, both for explanations which matched and mismatched engineers' problem-solving styles. 
We found that explanations were more frequently beneficial when they matched problem-solving style, but not every matching adaptation was equally beneficial; 
in some instances, diverse engineers found as much (or more) benefit from mismatched adaptations. 
Through an equity and inclusivity lens, our work highlights the benefits of having an LLM adapt its explanations to match engineers' diverse problem-solving style values, the potential harms when matched adaptations were not perceived well by engineers, and a comparison of how matching and mismatching LLM adaptations impacted diverse engineers.
\end{abstract}



\maketitle
\input{doc/01-Introduction}
\input{doc/02-background}
\input{doc/03-Methodology}

\input{doc/04_0_results_root}

\input{doc/07-Discussion-Limitations}
\input{doc/08-Conclusion}

\bibliographystyle{ACM-Reference-Format}
\bibliography{000-paper.bib}

\appendix

\newpage
\input{doc/09-Appendix}

\pagestyle{empty}
\end{document}

%% file: z-commands.tex

\definecolor{AbiOrange}{RGB}{251,212,180}
\definecolor{TimBlue}{RGB}{219,229,241}
\definecolor{WomanOrange}{RGB}{237,125,49}
\definecolor{ManBlue}{RGB}{31,78,121}
\definecolor{HelpfulBlue}{RGB}{2, 255, 255}
\definecolor{ProblematicYellow}{RGB}{255, 255, 84}


\newcommand{\briefCite}[1]{~\citeauthor{#1} (\citeyear{#1})}

\newcommand{\mysection}[1]{\section{#1}}
\newcommand{\mysubsection}[1]{\subsection{#1}}
\newcommand{\mysubsubsection}[1]{\subsubsection{#1}}

\newcommand{\colorboxBackgroundForegroundText}[3]%
{\protect\adjustbox{padding=1pt 1pt, bgcolor=#1}%
{\color{#2}#3}}%

\newcommand{\mapsize}{4cm}

\newcommand{\redact}[1]{
[Anonymized for Review]%
}%

\newcommand{\treatment}[1]{{%
#1}}

\newcommand{\feature}[1]{{%
#1}}

\newcommand{\quotateInset}[4]{%
    \vspace{0pt}%
    \begin{quote}%
    \user{#2}{ #3}{}{}: ``\textit{#1}''
    \end{quote}%
    \vspace{-2pt}%
}

\newcommand{\violationBox}[1]{
    \lfbox[border-style = dashed, border-color = red, border-width = 1px, padding = {0.5px, 1px, 0.5px, 1px}]{#1}%
}%

\newcommand{\vioProduct}[0]{
    {\fontfamily{cmtt}\selectfont Violation AI product}%
    }%

\newcommand{\appProduct}[0]{
    {\fontfamily{cmtt}\selectfont Application AI product}%
    }%

\newcommand{\applicationBox}[1]{
    \lfbox[border-color = red, border-width = 1px, padding = {0.5px, 1px, 0.5px, 1px}]{#1}%
}%

\newcommand{\quoteWithLLM}[7]{%
\vspace{-4pt}
\begin{quote}%
\personaVal{#1}{P#2}-#3-#4-\personaVal{#5}{#6}: ``\textit{#7}''%
\end{quote}%
}

\newcommand{\HAI}[1]{%
Human-AI Interaction}

\newcommand{\aiToHuman}[1]{%
Human$\leftarrow$AI}

\newcommand{\humanToAI}[1]{%
Human$\rightarrow$AI}

\newcommand{\llamaWithVal}[3]{%
\colorboxBackgroundForegroundText{#1}{black}{LLM-#2-#3}%
}

\newcommand{\fixfixfix}[1]{%
{\color{red} FIX FIX FIX---#1}}%

\newcommand{\probType}[1]{%
problem-solving style type}%

\newcommand{\probVal}[1]{%
problem-solving style value}%

\newcommand{\personaVal}[2]{%
\colorboxBackgroundForegroundText{#1}{black}{#2}}


\newcounter{boldifyCounter}
\newcounter{fixmeSectionCounter}
\newcounter{fixmeTotalCounter}
\newcounter{insightCounter}
\newcounter{insightSubCounter}
\newcounter{resultCounter}

\makeatletter
\@addtoreset{fixmeSectionCounter}{section}
\@addtoreset{fixmeSectionCounter}{subsection}
\@addtoreset{boldifyCounter}{section}
\@addtoreset{boldifyCounter}{subsection}

\makeatother

\newcommand{\boldify}[1]{}
\ifdefined\boldifyON
	\renewcommand{\boldify}[1]{\par\noindent%
		\textbf{{**}%
		 #1**}%
	}
\fi

\newcommand{\fakeResult}[1]{}
\ifdefined\fakeResultON
    \renewcommand{\fakeResult}[1]{\par\noindent%
        \stepcounter{boldifyCounter}%
        \textbf{{\color{blue}**FAKE FAKE FAKE**}%
        ~\arabic{section}.\arabic{subsection}.\arabic{boldifyCounter}%
        : #1\\}
    }
\fi

\newcommand{\FIXED}[1]{{\color{blue}#1}}
\newcommand{\FIXME}[2]{}
\ifdefined\fixmeON
	\renewcommand{\FIXME}[2]{\par\noindent%
		\stepcounter{fixmeSectionCounter}\stepcounter{fixmeTotalCounter}%
		{\color{red}\fbox{\color{black}%
			\parbox{.97\linewidth}{%
                \begin{center}
				\textbf{FIXME \arabic{section}.\arabic{subsection}.%
        		\arabic{fixmeSectionCounter} ({\color{red}%
        		\#\arabic{fixmeTotalCounter}})}
                
                \end{center}

                \textbf{#1: }#2
        		}
        		}%
        }
	}
\fi

\newcommand{\Result}[2]{\par\noindent%
    \stepcounter{resultCounter}
    \color{black}\fbox{\color{black}%
		\parbox{.97\linewidth}{%
             \textbf{Result \#\arabic{resultCounter}:}
             \textit{#1} #2
        }
    }
}

\newcommand{\llmSnippet}[3]{ \par\noindent%
   \stepcounter{insightCounter}
   {\color{black}
       \begin{tcolorbox}[%
            colback = white, 
            before = \par\noindent, 
            colframe = gray!50,
            top = 0pt,
            bottom = 0pt,
            boxsep = 0pt]
       \textbf{\textit{ #1 Explanation Snippet (#2 Program)} }
       \begin{quote}
            #3
       \end{quote}
       \end{tcolorbox}{\color{black}%
      }%
   }
}

\newcommand{\DP}[1]{\textcolor{blue}{[dp: #1]}}
\newcommand{\MMB}[1]{\textcolor{purple}{[mmb: #1]}}
\newcommand{\new}[1]{\textcolor{cyan}{#1}}
\newcommand{\AAA}[1]{\textcolor{orange}{[aaa: #1]}}
\newcommand{\JW}[1]{\textcolor{blue}{[jw: #1]}}

\newcommand{\note}[2]{}
\ifdefined\noteON
	\renewcommand{\note}[2]{\par\noindent%
		\stepcounter{fixmeSectionCounter}\stepcounter{fixmeTotalCounter}%
		{\color{blue}\fbox{\color{black}%
			\parbox{.97\linewidth}{%
                
                {
                \tiny
                \textbf{#1: }#2
        		} 
        		}%
          }%
        }
	}
\fi

\newenvironment{insight}[3]
    {\par%
    \setlength{\hangindent}{0.05\columnwidth}%
    \setlength{\parindent}{0.05\columnwidth}%
    \textit{Insight-G#1-#2: } #3\\%
    }%

\ifdefined\specialFooterON
\usepackage{fancyhdr, datetime}
\pagestyle{fancy}
\fancyhf{}
\pagenumbering{arabic}
\rfoot{\smallbreak{\tiny\today~\currenttime~GMT\quad\quad}Page~\thepage~}
\fi

\newcommand{\draftStatus}[2]{}
\ifdefined\draftStatusON
	\renewcommand{\draftStatus}[2]{
        \\ **#1 says D#2**%
	}%
\fi 

\newcommand{\doubleUnderline}[1]{
    \underline{\underline{#1}}
}

\newcommand{\programmingfont}[1]{%
    {
        #1%
}%
}%

\newcommand{\topic}[1]{} 
\renewcommand {\topic}[1]{%
     {\color{black}{#1}}%
    }

%% file: z-stat-commands.tex
\newcommand{\tTestResult}[5]{
    (#1-sided t-test, t(#2) = #3, \textit{p}: #4, \textit{d}: #5)%
}%

\newcommand{\pairedTResult}[3]{
    (pairedt-test, t(#1) = #2, \textit{p} = #3)%
}%

\newcommand{\oddsRatio}[3]{%
    (Odds Ratio: #1$, CI: [#2, #3])%
}%

\newcommand{\odds}[2]{%
    (Odds: #2)%
}%

%% file: doc/01-Introduction.tex
\section{Introduction
\draftStatus{AAA}{2.5}
}
\label{sec:intro}

\boldify{The problem: Can LLMs adapt their explanations to who they're talking to when explaining code? After all, people are diverse, and explanations aren't one-size-fits-all.
}

\topic{One problem in \HAI{} is whether and how large language models (LLMs) can adapt their explanations of code to fit the needs of diverse software engineers.}
Software engineers are diverse---they have diverse ages, gender identities, diverse races and ethnicities, diverse cultures, and more.
Prior works have shown that in other areas of Artificial Intelligence (AI), there may not be ``one-size-fits-all'' explanations~\cite{anderson2019explaining,anderson2020mental,dodge2021no} for diverse individuals.
To address this problem, we investigate the impact of an LLM adapting its code explanations to diverse software engineers---specifically for their diverse problem-solving styles.

\boldify{the area of problem-solving. Here's why this is problem-solving, and we know that there's room to adapt because people have diverse problem-solving styles.}

\topic{Why problem-solving styles?}
When people try to solve problems, they may seek additional information, which is exactly what LLM-generated explanations offer.
Recent research in \HAI{} has shown that users of AI products, like users of many other systems, have diverse approaches to problem-solving~\cite{anderson-diversity-2024,nam2024using}.
Our investigation draws specific problem-solving style types from the Gender Inclusiveness Magnifier (GenderMag), an inclusive design method with a dual problem-solving/gender focus~\cite{burnett2016gendermag}.
GenderMag identifies five different problem-solving style types: learning style, technological self-efficacy, attitudes toward risk, information processing style, and motivations for using technology.
Each of these types has a wide range of possible values, explained in detail in Section~\ref{sec:background-rw}.

In our study, we generated code explanations that adapted to the extreme values on each problem-solving style type.
We recruited 53 professional software engineers and showed each engineer three explanations---one which had not adapted to \textit{any} problem-solving style value, one that adapted to their \textit{same} problem-solving value for one type, and one that adapted to the \textit{opposite} value of that same type.
Engineers rated how they felt about each explanation and assessed the extent to which they thought those explanations helped them and fit their needs.



Given the amount of variability in skill and programming language familiarity amongst software engineers, we designed code explanation scenarios where prior knowledge would not play a major factor in their understanding of the source code. 
As such, we used source code written in the COBOL language, a language commonly used in mainframe computing applications\footnote{In 2016, David Powner~\cite{powner2016federal} testified to the U.S. House of Representatives that parts of government systems still ran on COBOL for the Department of Homeland Security, Department of Veteran's Affairs, and the Social Security Administration.
The following year, Reuters~\cite{reuters2017cobolblues} reported  significant usage of COBOL in the finance sector; for example, 95\% of ATM swipes still rely on COBOL code. Thus, it is important for LLM-based AI code assistants to be able to generate understandable explanations of COBOL code.} and one with which software engineers are not generally familiar~\cite{stackoverflow2024developer}.


\boldify{We apply two lenses to these data---an equity and inclusivity lens. Here's what we mean by equity and inclusivity.}

Our results reveal that when an LLM tried to adapt its code explanations to five different problem-solving style types, the adaptations' efficacies differed across the five types.
One measurable difference was inclusivity and equity of LLM adaptations, where inclusivity is measured within a single group of engineers between the unadapted LLM and its adaptations.
If engineers' were better supported by an LLM adaptation, we consider that adaptation as \textit{more inclusive}, relative to the unadapted.
Similarly, equity is measured between two groups of engineers, where if engineer group A were better supported than engineer group B for a given LLM, we consider that LLM as inequitable against group B.


\boldify{Our work makes the following contributions:}

Our work makes the following contributions to the \HAI{} community:

\begin{itemize}
    \item An empirical assessment of adapting LLMs' explanations to five diverse problem-solving styles from an existing inclusive design method for engineers with diverse problem-solving styles.
    \item An understanding of how adapted LLMs' explanations that match engineers' problem-solving style values can change inclusivity and equity for diverse engineers, but not \textit{always} for the better.
    \item A comparative analysis of the impacts of adapted LLMs' explanations that match engineers' \textit{opposing} problem-solving style value and their negative influences on inclusivity and equity for diverse engineers. 
\end{itemize}


%% file: doc/02-background.tex
\section{Background \& Related Work 
\draftStatus{AAA}{0}
}
\label{sec:background-rw}

In this section, we provide an overview of four areas relevant to our study: 1) the GenderMag method with relevant findings outside \HAI{} 2) \HAI{} which have taken an inclusivity lens to their research, 3) research on adapting LLMs, and 4) research on the studying human-LLM interaction.

\FIXME{AAA}{This stuff below is pulled verbatim from the inclusivity paper.
It'll have to change to avoid self-plagiarising }

\input{tables/01-GenderMag-Persona-Table}

\subsection{GenderMag: The Gender Inclusiveness Magnifier}
\label{sec:GenderMag}

\new{\boldify{The GM explained. Description of the Personas and the five facets.}}
The GenderMag method is an inclusive design and evaluation method and has a dual gender/problem-solving style focus~\cite{burnett2016gendermag}.
Software professionals use the GenderMag method to detect ``gender-inclusivity bugs''.
GenderMag empowers professionals in finding these issues \textit{not} by using people's gender identities but instead through five problem-solving style types, shown in Table~\ref{tab:01-GenderMag-Persona-Table}
GenderMag's dual focus comes from the literature repeatedly identifying the five problem-solving style types as having strong ties to both problem-solving and gender identity.
GenderMag's problem-solving styles fit well with studying AI technologies, becauseGenderMag's origins lie in improving \textit{inclusiveness} for \textit{problem-solving technology}---such as spreadsheet development, software debugging, or any other domain where people solve problemse~\cite{burnett2016gendermag}.
Consuming explanations, an area of focus in this paper, is inherently problem-solving---one reason people seek explanation is when there is a knowledge gap (problem) that they wish to fill (solving).

The five problem-solving style types in this paper each have a continuous range of values, where the endpoints (Table~\ref{tab:01-GenderMag-Persona-Table}'s column~2 \& 4) are distinguished values.
Values on one end of each type are assigned to a persona named ``Abi,'' and the other end's values are assigned to a persona named ``Tim.''
``Pat,'' the persona in the middle, catches a mix of values between ``Abi'' and ``Tim.'' 
For example, Abi prefers to learn by process when learning new technologies (Table~\ref{tab:01-GenderMag-Persona-Table}, row 1), leveraging tutorials, wizards, and how-tos.
However, Pat and Tim both prefer to learn by tinkering, constructing their own understanding of technology by clicking, exploring, and investigating;
what distinguishes Pat and Tim for learning style is hat Tim can sometimes tinker to excess, whereas Pat tinkers reflectively (e.g., considers what led them there and what they found along the way).
As such, Abi might follow a more structured approach, leveraging a ``recipe'' of sorts.

\boldify{Outisde of HAI, these five styles have been shown to be about problem-solving (with a wee bit of gender in the middle), such as learning style...}

Outside of \HAI{}, GenderMag's five styles have repeatedly been shown to be about \textit{problem-solving}.
We provide a brief description of each style here (see \citet{burnett2016gendermag} for more).
Learning style (Table~\ref{tab:01-GenderMag-Persona-Table}, row 1) examines how people structure their approaches to solving problems.
Some people (like ``Abi'' in the table) may take a methodical approach to solving problems, whereas others (like ``Tim'') prefer to learn by tinkering.
Organizations have started developing support for these diverse learning style approaches (e.g.,~\cite{kim-microsoft2023inclusive}).

\boldify{...and self-efficacy...}

People's technology self-efficacy (Table~\ref{tab:01-GenderMag-Persona-Table}, row 2) reflects one form of confidence---their beliefs that they can succeed at specific tasks~\cite{bandura1986explanatory}.
Self-efficacy influences problem-solving, since it influences 1) which cognitive strategies people apply to problems, 2) how much they persist when they encounter difficulties, and 3) which coping strategies they apply in the face of such difficulties~\cite{bandura1986explanatory,bandura1977self}.
A host of literature has tied diverse self-efficacies and approaches to problem-solving tasks across myriad domains (e.g.,~\cite{yaugci2016effect,padala2020gender,yang2023self,voica2020motivation}).

\boldify{...and risk...}

When solving problems, people have diverse attitudes toward risk (Table~\ref{tab:01-GenderMag-Persona-Table}, row 3).
Risk can mean \textit{any} kind of risk, such as the risk of wasting time, failing at a task, or losing control over their data used to train AI.
While engaging with technology, some people may be more risk-averse than others, where potential cost/benefit analyses might place more emphasis on the potential costs outweighing the benefits.
On the other hand, some people may place more emphasis on the potential benefits, more tolerant of potential associated risks~\cite{weber2002domain,charness2012strong,hyll2015impact,weber2002domain,charness2012strong}.

\boldify{...and IP...}

When people solve problems, they often require information to solve them, yet people gather information in diverse ways (Table~\ref{tab:01-GenderMag-Persona-Table}, row 4).
Some may wish to form a complete understanding of a problem before acting, gathering information more comprehensively before acting upon the information in a batch of activity.
Others may wish to gather information selectively, acting upon a first promising piece of information, pursuing that, and possibly gathering a little more after backtracking if the first piece does not work.
Those who investigate technology usage have paid attention to how technology empowers diverse information processors in obtaining the right \textit{amount} of information at the right \textit{time} (e.g.,~\cite{chapman2022USDSguidelines,pirollicard1995IFT,torrens2020lacking,whittaker2011personalInfoMgt}).

\boldify{...and motivations.}

The last problem-solving style considers \textit{why} people interact with technology---their motivations (Table~\ref{tab:01-GenderMag-Persona-Table}, row 5).
Some people may interact with technology mainly to accomplish some task, whereas others may interact with it because it is a source of fun and enjoyment for them~\cite{stafford2001identifying}.
People's diverse motivations can impact not only which parts of technology they decide to use but also how they use them~\cite{grosso2021exploring,stafford2001identifying,bridges2018hedonic,o2010influence,kartal2022preservice}.

What distinguishes this paper is that the context of most GenderMag findings currently exist outside of HAI~\cite{agarwal2023MOSIP, cunningham2016supporting,marsden2016evaluation,russian2017gender,shekhar2018cognitive,gralha2020genderDiffs,carver2018gender, padala2020gender,vorvoreanu2019gender,murphy2024gendermag,stumpf2020gender}.


\boldify{Our paper deals with fairness, and there are myriad papers on \textit{algorithmic} fairness; ours is NOT about algorithmic but instead on experiential fairness.}

\topic{In \HAI{} research, equity and inclusion commonly arise regarding algorithmic fairness.}
These approaches focus on detecting and fixing algorithmic decision making harms and have myriad works published (e.g.,~\cite{bird2020fairlearn,katzman2023taxonomizing,propublica-compas,harrison2020empirical,green2019disparate,buolamwini2018gender,yang2020towards}).
Instead, this work focuses on promoting fairness in user experience by accounting for the diversity of problem-solving styles that people have.

\boldify{The second area related to our paper is within the space of HAI, but they do not deal with LLMs.}

\topic{The field of \HAI{} has started to investigatea user-experience-centric approach to inclusivity and equity for diverse AI product users.}
One approach that researchers have taken involves looking at how people's gender identities can influence how they perceived how fair an AI's decisions were~\cite{van2021effect}, how likeable people found a product~\cite{derrick2014affective}, and even how people thought an AI reasoned~\cite{joseph2024artificial}.
However, other works have emerged which have started to consider the five problem-solving styles from Table~\ref{tab:01-GenderMag-Persona-Table}.
For instance, Kulesza et al.~\cite{kulesza2012tell} measured a change in participants' computer self-efficacy, given a ``why''-oriented explanation approach, and Jiang et al.~\cite{jiang2000persuasive} found that 
 particpants with higher self-confidence were less likely to accept an AI product's recommended solution.
Other researchers have found similar differences for attitudes toward risk~\cite{schmidt2020calibrating,cohen2020sensitivity}, information processing style and learning style~\cite{zhang2021information}, and motivations~\cite{shao2021hello,li2020understanding}.
One HAI paper that investigated \textit{all five} of these problem-solving styles was Anderson et al.~\cite{anderson-diversity-2024}, who found inclusivity and equity differences between when one of \citet{amershi2019guidelines}'s guidelines were violated or applied.
Our work differs from Anderson et al.~\cite{anderson-diversity-2024} by investigating \textit{real} AI outputs (i.e., LLM explanations), as opposed to vignettes, and by isolating all 10 of the problem-solving style values.

\boldify{One of three major areas of LLM research relevant to our own are the works who measure whether and how LLMs adapt their responses (but they don't have a user study)}

This paper studies both how LLMs adapt when provided with different prompts and an empirical evaluation of people interacting with LLMs.
Research focusing on LLM adaptations (without a user study) have considered multiple dimensions of adaptation to provide to an LLM, such as demographic diversity.
One instance is \citet{rooein2023know}, who prompted four LLMs to adapt to age and education levels.
Our work differs by prompting LLMs to adapt to problem-solving style, in an attempt to cut across demographic dimensions to inclusively adapt LLMs' code explanations.

\boldify{A second of these three areas deals with studying diverse people's interactions with an unadapted LLM (i.e., it hasn't been told to cater its explanations)}

The second part of this area has focused on studying how (and why) diverse people interact with an LLM (without adapting the LLMs' responses).
\cite{skjuve2024people} derived six motivations for why people interact with an LLM---ChatGPT---though their definition of  motivations was not the same as the motivations problem-solving style presented in this paper.
Some of the six motivations they uncovered were engaging with an LLM for productivity, novelty, creative work, and learning.
More domain-specific to our work with code explanations, Feldman \& Anderson~\cite{feldman2024non} investigated how diverse programming expertise (ranging from non-programmers--beginners) influenced perceptions and attitudes of an LLM.
They found that non-programmers struggled to interact with an LLM designed to provide code, and they were also more likely to blame themselves when something went wrong with the LLM than beginner programmers were.
Others have also reported similar findings with code expertise, finding that experience with programming influences not only whether and how diverse programmers use an LLM but also their perceptions of the LLM~\cite{chen2024learning,nguyen2024beginning} 

Particularly close to our work are those which have started considering some of the same problem-solving styles.
For example, Choudhuri et al.~\cite{choudhuri2024howfar} found that using an LLM could directly impact their participants' self-efficacies.
Nam et al.~\cite{nam2024using} investigated how software engineers' information processing and learning styles, gathered using a similar problem-solving style survey, impacted how they used an in-IDE LLM (named GILT).
Their work found that engineers' diverse learning styles significantly influenced how they interacted with GILT.

\boldify{Our work differs by existing at the intersection of these areas, which isn't empty. However, others have studied other dimensions of human diversity, rather than problem-solving style}

One distinguishing factor about our work from those above is that it exists at the intersection, both adapting an LLM's responses to diverse engineers \textit{and} empirically evaluating their experiences with such adaptations.





%% file: tables/01-GenderMag-Persona-Table.tex
\begin{table}[]
    \centering
    \includegraphics[width =\linewidth]{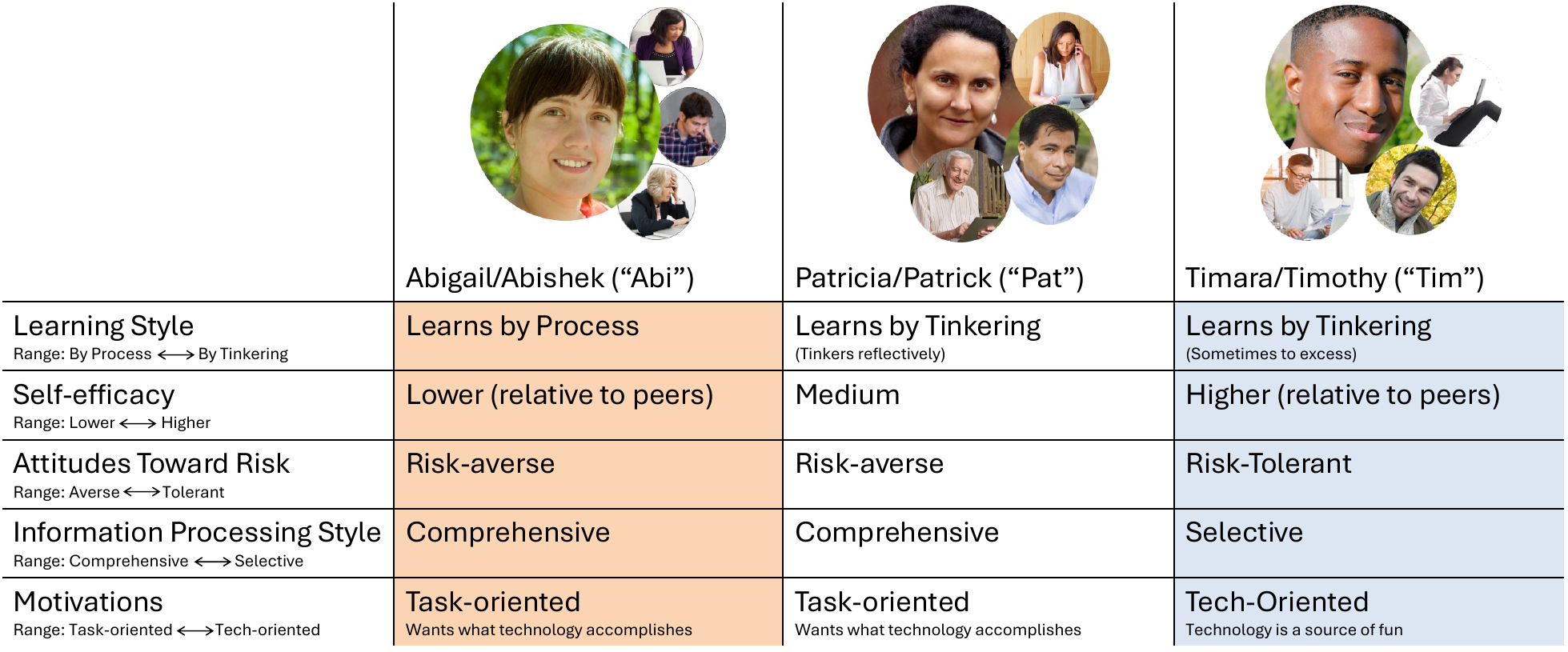}
    \caption{GenderMag's five problem-solving style types (row), and the values associated with each type (columns).
    Each of the five diverse problem-solving style types may have different needs from technologies.}
    \label{tab:01-GenderMag-Persona-Table}
\end{table}

%% file: doc/03-Methodology.tex
\section{Methodology
\draftStatus{AAA/DP}{2 (DONE)}}
\label{sec:methods}

\subsection{Research Questions}
\label{subsec:research_questions}

In this study, we seek to answer the following research questions:

\begin{itemize}[ label = {}, leftmargin= 0.5cm]
    

    \item \textbf{RQ1}---
    How equitably did code explanations serve the diverse engineers, when produced by \textit{unadapted} LLMs?
    

    \item \textbf{RQ2}---When LLM adapted to \textit{match} the engineers' problem-solving style values:
    \begin{enumerate}[ label = {}, leftmargin = 0.5cm]
        \item \textbf{Inclusion}---How well (inclusively) did code explanations  serve the diverse engineers, when produced by LLMs that adapted to the engineers' \textit{same} problem-solving style value?
        \item \textbf{Equity}---
        How equitably did code explanations serve the diverse engineers, when produced by LLMs adapted to the engineers' \textit{same} problem-solving style value?
    \end{enumerate}

    \item \textbf{RQ3}---When the LLM adapted to the \textit{opposite} of the engineers' problem-solving style values:
    \begin{enumerate}[ label = {}, leftmargin = 0.5cm]
        \item \textbf{Inclusion}---How well (inclusively) did code explanations  serve the diverse engineers, when produced by LLMs adapted to the engineers' \textit{opposite} problem-solving style value?
        \item \textbf{Equity}---
        How equitably did code explanations  serve the diverse engineers, when produced by LLMs adapted to the engineers' \textit{opposite} problem-solving style value? 
    \end{enumerate}

    
\end{itemize}


To answer our research questions, we prompted an open-source LLM to explain COBOL code for each of the problem-solving style values. 
Details regarding the participants, LLM prompting, and experiment design follow.

\subsection{Participants
\draftStatus{dp}{2}}
\label{subsec:experiment-participants}

\input{tables/02-Participant-Demographics}

\boldify{We recruited 53 software engineers who worked for \redact{IBM}, and they came from all walks of life.}

\topic{We recruited 53 software engineers from a large, international technology company who had no COBOL experience for this study.}
Of the participants who disclosed their demographics, half were in the 25--34 age range, 57.4\% identified as a man, 40.4\% identified as woman, and 2.1\% identified as non-binary. Table~\ref{tab:02-Participant-Demographics} provides more detail.
Participants were sent the equivalent of U.S. \$25 for their participation.

\subsection{Experiment Design
\draftStatus{dp}{2}}
\label{subsec:experiment_preparation}

\FIXME{AAA}{For infinite loop, we don't care if the LLM couldn't explain it.
We had experienced engineers, and if they caught that it was an infinite loop and that the LLM didn't explain it, then how would that impact ratings?}



\subsubsection{Selecting the COBOL programs
\draftStatus{dp}{2.5}}
\label{subsubsec:cobol_selection}

\boldify{We selected COBOL programs from an intro to COBOL programming textbook, choosing problems with concepts that were not get too advanced but also that were not too simplistic. We sandboxed to reduce to 3.}

\topic{The software engineers had no COBOL experience, so we selected introductory COBOL problems to prevent making the task too difficult.}
We gathered 9 COBOL programs from Coughlan's~\cite{coughlan2014beginning} open-source repository and added a Fibonacci sequence generator because the sequence frequently appears in first-year CS courses%
\footnote{Coughlan: https://github.com/Apress/beg-cobol-for-programmers/tree/master\\
Fibonacci: http://progopedia.com/example/fibonacci/341/ \\All 10 programs are in the supplementary documents.}.
We piloted these programs with five internal software engineers to remove those which were too easy/hard, resulting in three programs---the Fibonacci sequence, an infinite loop, and ``rock-paper-scissors.''
Table~\ref{tab:05_cobol_programs} shows these three programs.

\input{tables/05_cobol_programs}

\boldify{We had to obfuscate these programs to try and minimize the impact of software engineers' experiences.}

\topic{The programs in the repository provided clues to their functionality, so we renamed each program identifier to ``MyProgram,'' renamed variables, and removed any comments.}
For example, in the Fibonacci sequence program (Table~\ref{tab:05_cobol_programs}, left), the fib1, fib2, fib3, and fibst variables were renamed to F1, F2, F3, and FST.
When needed, we also renamed any named calls, such as ``Loop'' as seen in (Table~\ref{tab:05_cobol_programs}, middle)
\footnote{
In Coughlan's original version, this was named ``InfiniteLoop'', which was an immediate tell}.

\subsubsection{Adapting the LLM for five problem-solving styles
\draftStatus{dp}{2}}
\label{subsubsec:adapting_LLMs}

\boldify{Part of establishing the \textit{how} LLMs adapted was to use \textit{multiple} LLMs (1 chat, 1 instruct, 1 code). Here are the parameters we chose and a high-level look (Figure~\ref{fig:01-Programming-Methodology-Overview}).}

We chose llama-3-70b-instruct because it has been used in multiple studies to gauge its adaptability to both personas and specific domains~\cite{chan2024scaling,siriwardhana2024domain} as well as its ability to generate code explanations~\cite{dubey2024llama}.
Llama-3's parameters were held constant with the exception of the following. 
We set the token output size to be between 10 and 1024 and set the model's temperature to 0.5.

\topic{We created a set of 11 LLM system prompts groups into three possible adaptations.}
The unadapted prompt (called Unadapted LLM throughout) behaved as llama-3 would naturally.
Five of the adaptations tried to fit an \personaVal{AbiOrange}{``Abi''-like} engineer.
The remaining five adaptations tried to fit a \personaVal{TimBlue}{``Tim''-like} engineer.
In this paper, we refer to all LLMs with the following convention: LLM-StyleType-StyleValue, where StyleType refers to the problem-solving style types (the rows of Table~\ref{tab:01-GenderMag-Persona-Table}, like learning style or technology self-efficacy) and StyleValue refers to values along each type (the cells on each row of Table~\ref{tab:01-GenderMag-Persona-Table}, such as \personaVal{AbiOrange}{process-oriented} or \personaVal{TimBlue}{higher} self-efficacy).
Figure~\ref{fig:system-prompt-viz} details how we operationalized the adaptations in the system prompt, providing a snippet for adapting llama-3 to a \personaVal{TimBlue}{risk-tolerant} engineer. 
The LLM was given an explicit mention of the problem solving style a set of statements that its user was more likely to agree/disagree with from the GenderMag problem-solving style survey~\cite{anderson-diversity-2024}

\bgroup
\def\arraystretch{1.5}%
\begin{figure}[t]
\small
\begin{tabular}{p{0.8in}p{4.8in}}
\toprule
\textbf{Prompt Section} & \textbf{Prompt Text} \\ \midrule
Instructions & You are a programmer's assistant. You can answer conceptual programming questions and explain what code samples do. You customize your responses to fit what you think you know about the person who has prompted you. Your responses are helpful and harmless and should follow ethical guidelines and promote positive behavior. Your responses should not include unethical, racist, sexist, toxic, dangerous, or illegal content. Ensure that your responses are socially unbiased. When responding to your user, you should talk directly to them. For example , responses like 'your user has a certain trait' should instead say 'you have a certain trait.' \\
Problem-solving style description & Research has shown that people have diverse attitudes toward risk, and risk can mean ANY risk, such as the risk of losing privacy, wasting time, or losing control of an AI. Additionally, research also suggests that when using AI technology, a user's experience with AI can differ by their specific risk attitude. \\
Problem-solving style details & Your user has a more risk-tolerant attitude than their peers, meaning they are more likely to disagree with the following statements:
\begin{enumerate}
    \item I avoid using new apps or technology before they are well-tested.
    \item ...
\end{enumerate}

Additionally, your user is more likely to agree with the following statements:
\begin{enumerate}
    \item I am not cautious about using technology.
    \item ...
\end{enumerate} \\
  Request and code & Can you explain what the following COBOL code does? ... \\ \bottomrule
\end{tabular}
\caption{Portion of the LLM prompt used for the \personaVal{TimBlue}{``Tim''-like} attitude towards risk. Each adaptation's prompt was composed using the same four sections (appended together into a single prompt), with the problem-solving style description and details changed to suit each \personaVal{TimBlue}{``Tim''-like} or \personaVal{AbiOrange}{``Abi''-like} problem-solving styles. The '...' indicate text that has been removed for this figure for clarity.}
\label{fig:system-prompt-viz}
\normalsize
\end{figure}
\egroup

\boldify{We looked at each of the three LLMs' outputs, and we selected Llama-3-70b-instruct not only because it seemed to adapt more readily to \textit{all} of the different prompts but also because its effectiveness has been empirically demonstrated~\cite{castricato2024persona}.}

We asked each llama-3 adaptation to explain three COBOL programs each.
We did not prompt engineer, instead using readily available materials from the literature regarding these problem-solving style values.
The LLMs generated a total of 33 outputs, representing the 11 possible prompts given to each of the three LLMs that we tested.
Participants were not told the model that we used to prevent participants from bringing in any product biases to the study.
Instead, they were only ever referred to as Adastra, Boreas, and Caelum,

\subsection{Study Procedure
\draftStatus{dp}{2.5}}
\label{subsec:participants_did}

\boldify{We created medium fidelity slide decks that reflected each of the possible combinations of problem-solving style value/COBOL Program/LLM order, and here's why we chose medium fidelity}

\topic{Participants needed to interact with each LLM adaptation's response, for every combination of $\{problem-solving\ style\} \times \{COBOL\ Program\} \times \{Adaptation\}$.}
Thus, we created medium fidelity slide decks for each participant's treatment (like in Figure~\ref{fig:03_prototype_info_proc_boreas_infinite}).
We chose to use low/medium-fidelity prototypes, rather than interactive systems with the ability to directly prompt the LLMs, for the following reasons:
(1) This level of prototyping has shown comparable levels of success in user studies when compared against high-fidelity~\cite{walker2002high}.
(2) It prevented participants from prompting each LLM adaptation.
(3) Since participants knew that they were interacting with a slide deck, we thought that they would be more focused on LLM responses (e.g., content, structure, and missing explanations) as opposed to font size, colors, and other extraneous details found in a high-fidelity prototype.

\input{figures/03_prototype_info_proc_boreas_infinite}

\boldify{Before coming to a session, participants filled out the GenderMag problem-solving style survey found in Anderson et al.~\cite{anderson-diversity-2024}, and their five problem-solving style values were classified relative to their peers}

\topic{Prior to running any session, the first 30 participants filled out the GenderMag problem-solving style survey from Anderson et al.~\cite{anderson-diversity-2024}.}
This enabled us to classify each participant's five problem-solving style values using a median split, relative to the remaining participants in the study.
Since the GenderMag classification rules (Table~\ref{tab:06_gendermag_rules}) for each of the problem-solving style types rely on the median (robust against outliers) it was reasonable to assume that even new participants signing up would not change the problem-solving style values of finished participants%
\footnote{This remained true at the conclusion of the study. 
We formulaically checked that \textit{every} participant's original problem-solving style value was identical to their original assignment upon recruitment.}.
Once the first 30 participants had filled out the survey, we derived their problem-solving styles values and scheduled them.

\FIXME{MMB}{Not clear to reader what the overall expt design is (admittedly, this is after a skim, so I may have missed it, but I expected it to be in THIS subsec or at the top of methodology): \\
(1) why just 30? eg, is there a "control" group of about 30 where you don't care what their problem-solving styles is? \\
(2) later in results, there's a picture with a blue bar and an orange bar, but I'm not sure what that implies about the expt.  Are the blue the folks who said they were "Tim"s and the orange "Abi"s according to one of the problem-solving styles? If so, where are the ones who didn't fill out the questionnaire}

\input{tables/06_gendermag_rules}

\boldify{We randomly assigned participants' treatment (the problem-solving style type) and which of the rows in the Latin Square they would see.
The first author moderated \textit{every} session, and there were some housekeeping things to do.}

\topic{Before each participant's session, we randomly assigned them to one of the five problem-solving style treatments (between-subjects), and each participant randomly saw all three COBOL programs, explained by a different llama-3 adaptation (within-subjects)\footnote{For details on how participants were randomized, please see the Appendix.}.} 
Two llama-3 adaptations each adapted to an endpoint of the assigned problem-solving style type treatment, and the last was a control (no adaptations).

The first author moderated all sessions, where each participant was first asked to perform a think-aloud warm-up exercise to encourage them to talk throughout the session.
After, they were exposed to a COBOL ``Hello, World!'' program without \textit{any} LLM output to encourage participants' curiosity;
by exposing them to COBOL syntax without explanation or the ability to seek answers, we hypothesized that participants would have additional motivation to seek answers in the LLM explanations in the main task.

\boldify{Participants led the discussion during the main task of each session, where they thought aloud about what they found Helpful/Problematic for 15 minutes/LLM, highlighting \textit{anything} that they found helpful or problematic for them.}

\topic{Participants were given 45 minutes for the main task, where participants led the discussion on what they thought of the LLMs' explanations to answer our research questions.}
Before the task started, participants were encouraged to consider three topics of information---the content (words the LLM chose), the structure (the format the LLM chose), as well as what the LLM failed to explain that the participant wish they had.
Each participant had 15 minutes per LLM, where they highlighted and talked about which parts of the explanation they found helpful or problematic and \textit{why} they found it helpful/problematic.

\boldify{During these 15 minutes, participants also answered a battery of questions, pulled from the literature ( Feelings~\cite{benedek2002measuring}, perceived usefulness~\cite{reichheld2011ultimate}, trust~\cite{jian2000foundations},~\cite{miehling2024language}, and three applicable TLX dimensions~\cite{hart2006nasa}.}

\topic{During each 15 minute block with the LLMs, participants rated their experiences with each explanation.}
We used previously-published scales from \citet{li-MSR-work} to assess feelings, \citet{jian2000foundations} to assess trust, and \citet{hart1988development} to assess cognitive load. 
Confirmatory factor analyses confirmed a high degree of reliability to these scales (feelings: Cronbach’s $\alpha = .75$; trust: Cronbach’s $\alpha = .81$\footnote{During analysis, we observed that dropping a single item from this scale, ``not harmful,’’ resulted in an increased reliability from .75 to .81. Therefore, we opted to proceed with a modified trust scale by dropping this item.}; cognitive load: Cronbach’s $\alpha = .84$).
We also used the perceived usefulness question from \citet{davis1989technology}'s technology acceptance model (TAM) and conversational maxims with LLMs from \citet{miehling2024language}.
We identified a new factor through exploratory factor analysis comprised of helped, perceived useful, enough info, and well-organized.
We labeled this factor as ``utility’’ and noted that it was highly reliable (Cronbach’s $\alpha = .91$).
Questions which did not load into any factor or fell below the 0.7 threshold were excluded from analyses.

\input{tables/10_llm_dv_table}

\subsection{Data Analyses \& Validation
\draftStatus{dp}{2}}

\boldify{We have two equity-related research questions, and we look at equity between-subjects, specifically looking for 10\% gaps.}

\topic{To answer the equity-related research questions, we looked at differences in engineers' average ratings between two \probVal{}s.}
For example, to answer \textbf{RQ1} for the learning style \probType{}, we compared $|avg(process\ oriented) - avg(tinkering\ oriented)|$ for each dependent variable.
We compared this distance to a bound of 10\%, the equivalent of a letter grade difference in US grading (i.e., a ``B'' vs. a ``C'').
In general, between two opposing \textit{\probVal{}s} from Table~\ref{tab:01-GenderMag-Persona-Table}, a dependent variable as \textit{in}equitable if the distance between averages was at least 10\% of the total response range.

\boldify{We answer inclusivity-related questions within-subjects, we looked at 5\% differences \textit{within} problem-solving style values}

\topic{To answer the inclusivity-related research questions, we looked at differences in engineers' average ratings within the \textit{same} \probVal{} between two LLMs.}
For example, for the process-oriented \probVal{}, we compared $|avg(adapted) - avg(unadapted)|$ for each dependent variable, to get a sense for how much the average ratings changed between two explanations.
We again compared this distance to the same 10\% bound as the equity-related research questions.
Generally, within the same \textit{problem-solving style} (but between two LLMs), a dependent variable had increased or decreased inclusivity if the distance between averages was at least 10\% of the total response range.

%% file: tables/02-Participant-Demographics.tex
\begin{table}[b]
    \centering
    \begin{tabular}{c|cc|cc|c}
        \toprule
         \multicolumn{2}{c}{\textbf{Age Range}} & \multicolumn{2}{c}{\textbf{Gender Identity}}  & \multicolumn{2}{c}{\textbf{Work Exp.}}\\
         \midrule         
         18--24 & 10 & Woman        & 19    & 0--5  & 34 \\
         25--34 & 25 & Man          & 27    & 6--9  & 10 \\
         35--44 & 8 & Non-Binary   & 1     & 10+   & 9\\
         45--64 & 7 & &  & & \\
         \hline
         \textbf{Total:} & 50 & \textbf{Total:} & 47 &  \textbf{Total:} & 53\\
         \bottomrule

    \end{tabular}
    \caption{Participants' demographics, showing their diverse ages, gender identities, and years of work experiences as developers.
    Not all participants disclosed their demographic information, so not all totals (bottom) sum to 53.}
    \label{tab:02-Participant-Demographics}
\end{table}

%% file: tables/05_cobol_programs.tex
\begin{table}[t]
    \centering
    \includegraphics[width=0.95\linewidth]{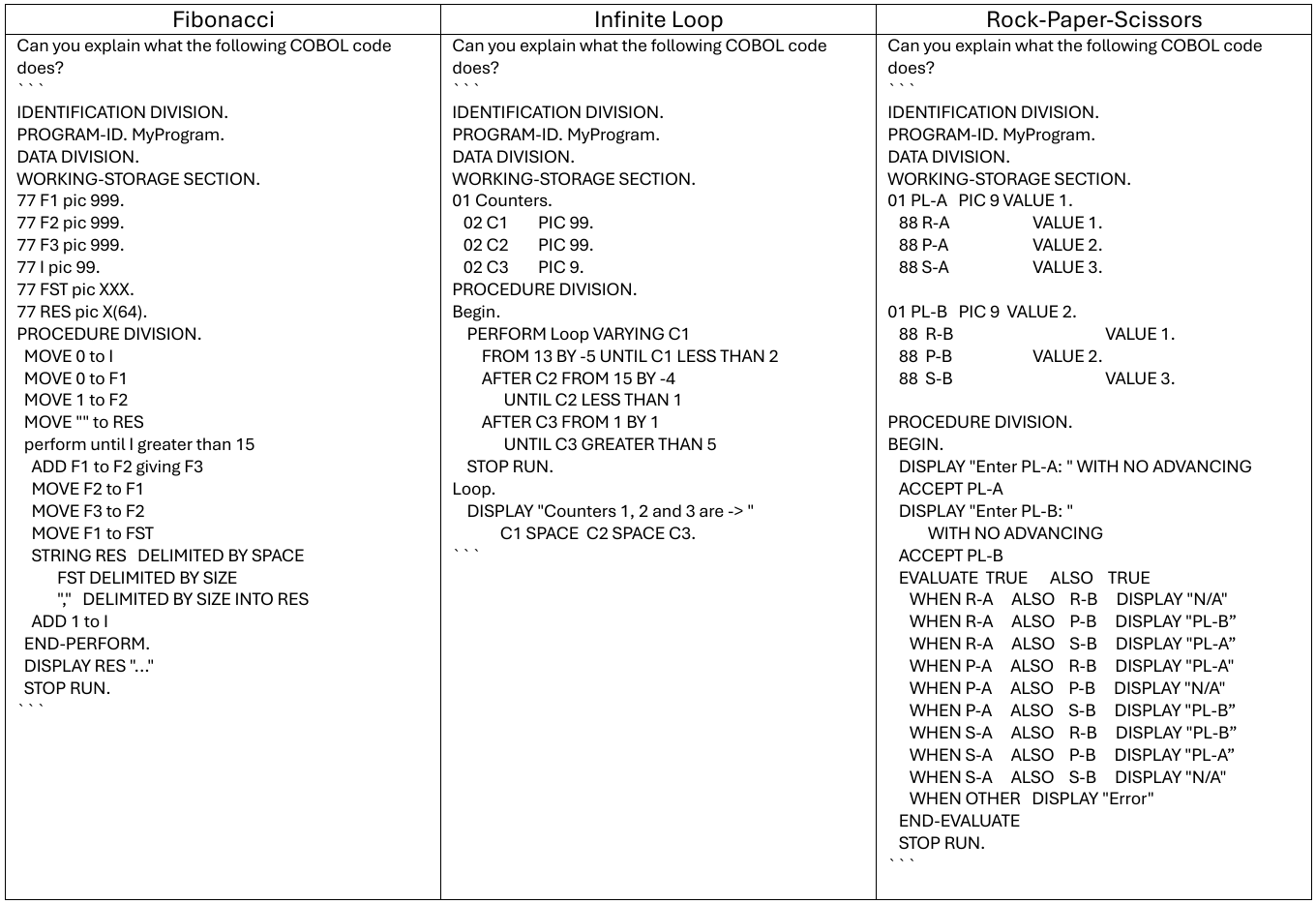}
    \caption{The three experiment COBOL programs.
    To encourage engineers to engage with the LLM explanations, variables were obfuscated (e.g., Fib1 $\rightarrow$ F1 in Fibonacci, Rock-A $\rightarrow$ R-A in Rock-Paper-Scissors).
    }
    \label{tab:05_cobol_programs}
\end{table}

%% file: figures/03_prototype_info_proc_boreas_infinite.tex
\begin{figure}[t]
    \centering
    \includegraphics[width=\linewidth]{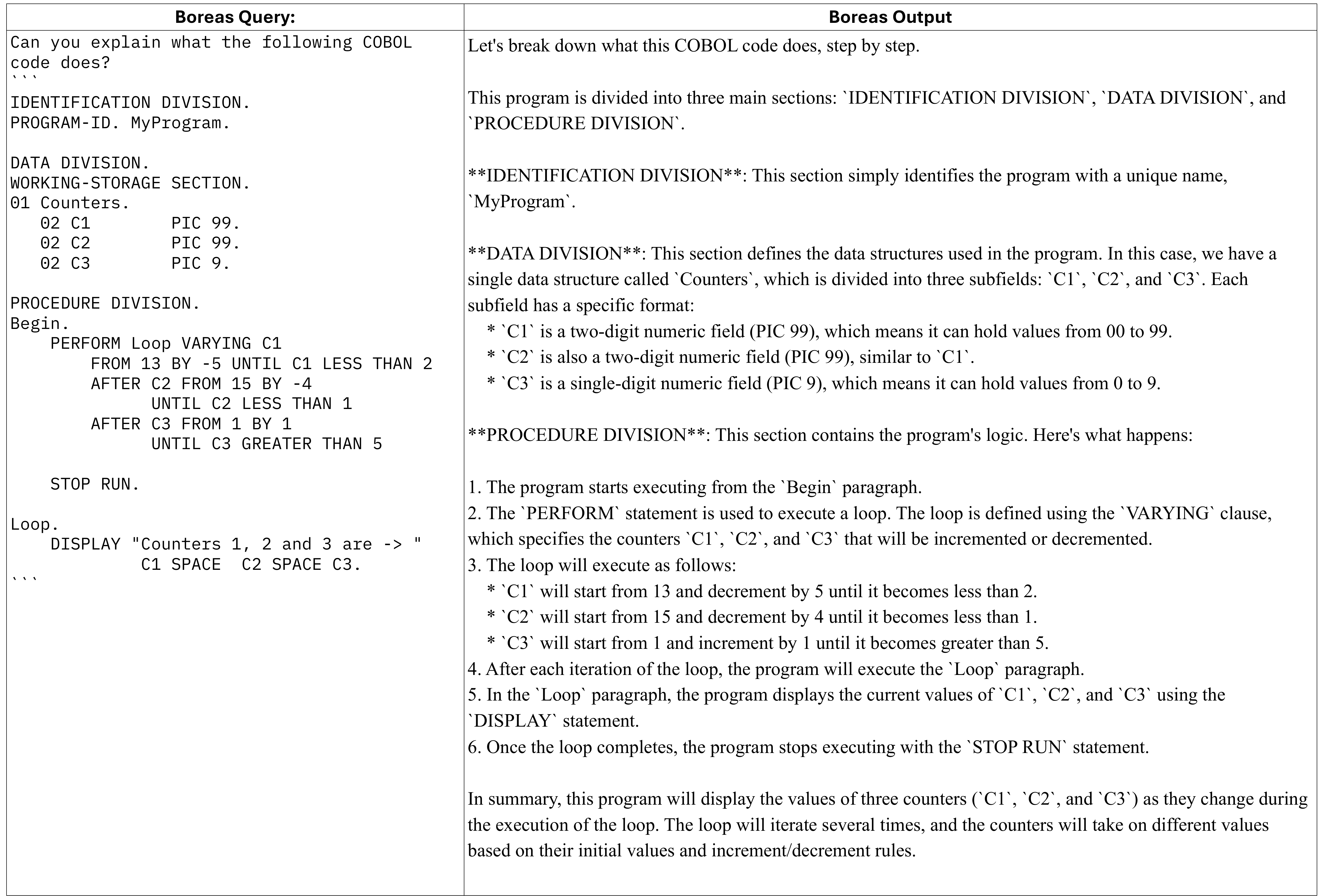}
    \caption{What participants saw in the information processing style treatment for llama-3's Boreas adaptation to the Infinite Loop COBOL program (Table~\ref{tab:05_cobol_programs}).
    We used medium fidelity prototypes for this study because of prototypes' comparable success to interactive systems in the literature and also to maintain participants' focus on what was helpful/problematic about \textit{these} outputs, rather than losing the participants to endless prompting or fixations on style/color/etc.}
    \label{fig:03_prototype_info_proc_boreas_infinite}
\end{figure}

%% file: tables/06_gendermag_rules.tex
\begin{table}[t]
    \centering
    \small
    \begin{tabular}{p{0.34\linewidth}|c|c|c}
        \toprule
        \textbf{Problem-Solving Style Type} & \textbf{Median Split Rule: If...} & \textbf{True} & \textbf{False}  \\
        \midrule
        Information Processing Style 
        & ...score $\geq $ median( scores ) & 
        \cellcolor{AbiOrange} Comprehensive & 
        \cellcolor{TimBlue} Selective \\
        Learning Style (by Process. vs. by Tinkering) & ...score $\geq$ median( scores ) & \cellcolor{AbiOrange} Process & \cellcolor{TimBlue} Tinker\\
        Self-efficacy & ...score $\leq$ median( scores ) & \cellcolor{AbiOrange} Lower & \cellcolor{TimBlue} Higher \\
        Attitudes Toward Risk & ...score $\geq$ median( scores ) & \cellcolor{AbiOrange} Averse & \cellcolor{TimBlue} Tolerant \\
        Motivations & ...score $\leq$ median( scores ) & \cellcolor{AbiOrange} Task & \cellcolor{TimBlue} Tech\\
        \bottomrule
    \end{tabular}
    \caption{GenderMag Problem-solving style value classification rules for each problem solving style type (rows).}
    \label{tab:06_gendermag_rules}
\end{table}

%% file: tables/10_llm_dv_table.tex
\begin{table}[]
    \centering
    \small
    \begin{tabular}{p{0.075\linewidth}p{0.125\linewidth}p{0.725\linewidth}}
        \toprule
         \textbf{Factor ($\alpha$)} & \textbf{DV Name} & \textbf{Wording} \\
         \midrule
         
         \multirow{5}
         {.08\linewidth}
         {Feeling ($.75$)}
         & In Control & ``I felt in control while interacting with [its] response''\\
         & Secure & ``I felt secure while interacting with [its] response''\\
          & Adequate$^\dagger$ &  ``I felt adequate while interacting with [its] response''\\
         &  Certain$^\dagger$ &   ``I felt certain while interacting with [its] response''\\
         & Productive & ``I felt productive while interacting with [its] response''\\
         \hline
         \multirow{4}{.08\linewidth}{Trust ($.78$)}& Not Suspicious$^\dagger$ &  ``I am not suspicious of [its]’s intent, action, or output.''\\
         & Not Harmful$^\dagger$ &  ``[its] responses will not have a harmful or injurious outcome.''\\
         & Unwary$^\dagger$ &  ``I am not wary of [its] response.''\\
         & Trust &  ``I can trust [its].''\\
         \hline
         \multirow{3}{.08\linewidth}{Cognitive Load ($.84$)} & Mental Demand & How mentally demanding was processing [its] response?\\
         & Performance & How successful were you in processing [its] response?\\
         & Effort & ``How hard did you have to work to accomplish your level of performance?''\\
         \hline
         \multirow{4}{\linewidth}{Utility ($.91$)} & Helped & ``[its] response helped me understand the COBOL program.''\\
         & Perceived Useful & ``I found [its] useful''\\
         & Enough Info & ``[its] response provided me with enough information.''\\
         & Well-Organized & ``[its] response was clear, unambiguous, and presented in a well-organized fashion.''\\
         \bottomrule
    \end{tabular}
    \caption{The 16 questions which participants rated in the study, grouped by their confirmatory/exploratory factor loadings.
    Feelings and Trust pulled from \citet{li-MSR-work}, \citet{jian2000foundations} respectively.
    Cognitive Load pulled from \citet{hart2006nasa}.
    Utility \& Quality questions pulled from both \citet{davis1989technology,miehling2024language}.
    $\dagger$ : Reverse-coded ratings.}
    \label{tab:10_llm_dv_table}
\end{table}

%% file: doc/04_0_results_root.tex
\section{Results: Were the LLM's unadapted explanations inequitable for diverse engineers? 
\draftStatus{AAA}{2.5+}
}
\label{sec:results_unadapted_equity}


\boldify{The answer to RQ1 was that yes, they were.}


\topic{We begin by reconsidering  whether there was even a need for the LLM to adapt to an engineer's problem-solving styles (\textbf{RQ1}).}
As Figure~\ref{fig:07_all_unadapted_LLM_inequities} shows, the answer was an emphatic yes---across the five treatments (y-axis), the engineers experienced 38 inequities in their responses (x-axis).
The \personaVal{AbiOrange}{``Abi''-like} engineers were particularly disadvantaged in three problem-solving styles---their Learning Style, Self-Efficacy, and Motivations---and the \personaVal{TimBlue}{``Tim''-like} engineers were particularly disadvantaged in one, namely their attitude toward Risk.
Finally, the LLM doled out disadvantages fairly evenly to both \personaVal{AbiOrange}{Abi-like} and \personaVal{TimBlue}{Tim-like} Information Processing Styles.

\input{figures/dv_equity_figures/07_all_unadapted_LLM_inequities}

\input{doc/04_1_results_rq_equity_inclusion_matching_vals}

\input{doc/04_2_results_rq_equity_inclusion_mismatching_vals}

%% file: figures/dv_equity_figures/07_all_unadapted_LLM_inequities.tex
\begin{figure}[h]
    \centering
    \includegraphics[width = .75\linewidth]{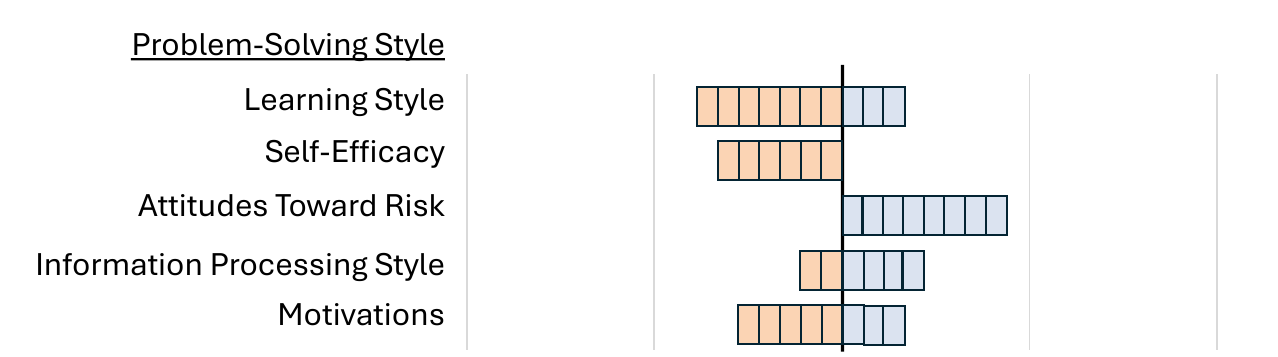}
    \caption{Unadapted LLM Inequities: The number of inequitable ratings for each problem-solving style type (centered at the black axis). 
    The LLMs were inequitable for all five problem-solving types, but the Orange (left of axis) \personaVal{AbiOrange}{``Abi''-like} problem-solvers were more frequently disadvantaged than the Blue (right of axis) \personaVal{TimBlue}{``Tim''-like} problem-solvers. }
    \label{fig:07_all_unadapted_LLM_inequities}
\end{figure}


%% file: doc/04_1_results_rq_equity_inclusion_matching_vals.tex
\section{Results: The LLM's Adaptation Matches---Relationships Between Inclusion \& Equity
\draftStatus{(Top) AAA}{2.5+}}
\label{sec:adapt_match}


\boldify{RQ2 asked about changes in inclusivity and resulting equity between the unadapted (baseline) and when llama-3's adaptations matched engineers' problem-solving style values. Three relationships between inclusivity and equity emerged from these data...}

\topic{\textbf{RQ2} asked how an LLM's adaptations that matched diverse engineers' problem-solving style values changed inclusivity and equity, relative to unadapted explanations.}
Table~\ref{tab:12_inclusive_equity_match_story_truth_table} provides the bottom line, showing for each problem-solving style type (rows) whether matching adaptations increased ($\uparrow$) or decreased ($\downarrow$) inclusivity for ``Abi''-like and ``Tim''-like engineers (middle columns).
The last column shows how the matching adaptations' inclusivity increased ($\uparrow$) or decreased ($\downarrow$) equity between the two groups.
Engineers' ratings showed three patterns in these data across the five problem-solving style types, discussed in more detail in each subsection:

\begin{enumerate}[leftmargin = 0.5cm, label = (\arabic*)]
    \item Increased inclusivity for both problem-solving style values increased equity (1 problem-solving style type).
    \item Increased inclusivity for one problem-solving style value increased equity (2 problem-solving style types).
    \item Decreased inclusivity for either problem-solving style value decreased equity (2 problem-solving style types).
\end{enumerate}
Sections~\ref{subsec:more_for_both_more_equity}--\ref{subsec:more_for_one_more_equity} provide one example of each pattern, and the supplemental docs provide all pattern examples.

\input{tables/12_inclusive_equity_match_story_truth_table}

\subsection{Pattern 1: Increased inclusivity for both problem-solving style values $\rightarrow$ increased equity
\draftStatus{AAA}{2.5}}
\label{subsec:more_for_both_more_equity}

\boldify{The first pattern was a ``best case'' senario for inclusivity and equity, where inclusivity supported ABi and Tim, and equity improved.}


\topic{The first pattern, which occurred for the learning style groups, was a ``best case'' scenario for inclusivity and equity (Table~\ref{tab:12_inclusive_equity_match_story_truth_table}, first row).}
In this pattern, the LLM's adaptations increased inclusivity for both the ``Abi''-like and the ``Tim''-like engineers' learning style values (a `$\uparrow$' in each cell).
Although inclusivity increases do not guarantee equity increases, the LLM's adaptations increased equity \textit{between} the two groups in this pattern (a `$\uparrow$' in the last column).

\boldify{Figure~\ref{fig:01_learn_inclusion_for_me} (left) shows just how inclusive the process-oriented adaptation was for the process-oriented engineers, helping in almost every variable in four categories. }

\topic{Figure~\ref{fig:01_learn_inclusion_for_me} (left) shows how often the LLM's process-oriented explanations (\llamaWithVal{AbiOrange}{Learn}{Process}) increased inclusivity for process-oriented engineers.}
The 12 boxes on the x-axis represent each rating where \llamaWithVal{white}{Learn}{Process} inclusively increased the process-oriented engineers' average ratings by 10+\% and decreased \textit{none} of their ratings by a 10+\% decrease.
\llamaWithVal{white}{Learn}{Process} also increase almost every rating's average in four categories (y-axis)---Feelings, Trust, and Utility. 

\input{figures/dv_inclusion_for_me_figures/01_learn_inclusion_for_me}


Table~\ref{tab:01_llm_learning_style_adapt_sections} shows an example of differences between the unadapted LLM (left) and the LLM adapted to process-style learners (right).  
As the figure shows, the adaptations were fairly subtle---in both versions, the LLM explained the code in a way that roughly followed the code structure.
However, the adaptation for process-style learners added additional information about the \textit{purpose} of the different sections, which was very helpful to process-oriented learners' ability to formulate a process for delving into the code.

\input{tables/LLM_prob_solve_adaptations/01_llm_learning_style_adapt_sections}


For example, as process-oriented participants P22 and P31 explained:%
\footnote{Note: Participant quotes attribute the PID-ProblemSolvingStyle-ProblemSolvingValue-llmSameValue/OppositeValue.\\
llmSameValue $\implies$ the LLM adapted to the engineers' same problem-solving style value.\\
OppositeValue $\implies$ the LLM adapted to the engineers' opposite problem-solving style value.}%
:
\quoteWithLLM{AbiOrange}{22}{Learn}{Process}{AbiOrange}{llmSameValue}{This is quite nice, how step-by-step this is.}
%
%
\quoteWithLLM{AbiOrange}{31}{Learn}{Process}{AbiOrange}{llmSameValue}{There were a few points where I was a little confused, but once I read the whole thing and jumped back, that was pretty solid...these headers [highlights the **IDENTIFICATION DIVISION** header] were helpful...to be able to jump up and down...}

\boldify{Similarly, when the \personaVal{TimBlue}{tinkering-oriented learners} saw \llamaWithVal{TimBlue}{Learn}{Tinker}, their ratings \textit{also} improved at a similar rate, with 12/19 ratings supported (63\%) and \textit{none} being undermined.}

\topic{\llamaWithVal{TimBlue}{Learn}{Tinker} provided a similar inclusivity win for \personaVal{TimBlue}{tinkering-oriented} learners.}
Returning to Figure~\ref{fig:01_learn_inclusion_for_me}, the right side shows that \llamaWithVal{white}{Learn}{Tinker} increased inclusivity for 11 of the tinkering-oriented learners ratings and never decreased them.

\boldify{As to equity, the inclusivity gain improved equity between the two groups (i.e., reduced differences)
}

\topic{Equity also fared well in this pattern.}
The left side of Figure~\ref{fig:01_learn_equity_changes_same_val} shows not only \textit{which} engineers (x-axis) the unadapted LLM had inequitably disadvantaged, but also \textit{where} these inequities arose across the four categories (y-axis).
And as the figure's right side shows,  
both \llamaWithVal{AbiOrange}{Learn}{Process} and \llamaWithVal{TimBlue}{Learn}{Tinker} improved equity, with fewer inequities disadvantaging either group of engineers.

\input{figures/dv_equity_changes_same_val/01_learn_equity_changes_same_val}


\input{figures/dv_inclusion_for_me_figures/02_se_inclusion_for_me}

\subsection{Pattern 2: Increased inclusivity for one problem-solving style value $\rightarrow$  increased equity
\draftStatus{AAA}{2.5+}}
\label{subsec:more_for_one_more_equity}

\boldify{The second pattern was still a ``good'' scenario for inclusivity and equity, where the adapted explanations increased inclusivity for  Abi OR Tim, and equity still increased.}

\topic{The second pattern was also a win for both inclusivity and equity (recall Table~\ref{tab:12_inclusive_equity_match_story_truth_table}, rows 2 \& 3).}
In this pattern, the LLM's adaptations increased inclusivity for either the ``Abi''-like or ``Tim''-like engineers' values, but not both (one `$\uparrow$' in the middle columns).
Here, the one-sided increase in inclusivity was for the very group that had been disadvantaged in the unadapted (inequitable) LLM, so the increase in inclusivity brought an increase in equity as well (`$\uparrow$' in the last column). 
This pattern occurred for both self-efficacy and risk; here we present the self-efficacy findings.  
(The risk findings can be found in the Supplemental Documents.)



\boldify{In other treatments, inequities in the engineers' ratings of llama-3's unadapted explanations were more skewed. Figure~\ref{fig:03_se_risk_unadapted_equity} shows that the ``Abi''-like risk-averse (left) and ``Tim''-like higher self-efficacy (right) engineers were inequitably advantaged.}

\topic{Figure~\ref{fig:02_se_inclusion_for_me} shows how much the LLM's adaptations to the engineers' computer self-efficacy values changed inclusivity.}
The left shows that \llamaWithVal{AbiOrange}{SelfEff}{Lower} increased inclusivity for six of the \personaVal{AbiOrange}{lower} self-efficacy engineers' ratings (left) in the Feelings, Trust, and Cognitive Load categories.
However, as the right side shows, the adaptions intended for \personaVal{TimBlue}{higher} self-efficacy engineers did not make much difference for those engineers.

Table~\ref{tab:02_llm_se_adapt_supportive_language} shows three versions of the LLMs' code explanations.
The unadapted version (top) jumped into the technical details, and \llamaWithVal{TimBlue}{SelfEff}{Higher} added little to this (bottom). 
In contrast, \llamaWithVal{AbiOrange}{SelfEff}{Lower} (middle) inserts supportive language, setting an expectation for the COBOL code's understandability and that the LLM will help the engineer.

\input{tables/LLM_prob_solve_adaptations/02_llm_se_adapt_supportive_language}

\boldify{For the lower self-efficacy engineers, who had more frequent inclusivity increases, sometimes made statements that reflected what they liked about it.}

The engineers' ratings in Figure~\ref{fig:02_se_inclusion_for_me} made it clear that the \personaVal{AbiOrange}{lower} self-efficacy engineers seemed to appreciate \llamaWithVal{AbiOrange}{SelfEff}{Lower}'s approach.
As P14 put it:
\quoteWithLLM{AbiOrange}{14}{SelfEff}{Lower}{AbiOrange}{llmSameValue}{This is trying to reassure me with some language.
I think this is a good thing overall.
... a little bit more humanity out of a response like this.
}

\boldify{Success! Similar to Section~\ref{subsec:two_inclusivities_resolves_inequity}, these inclusivity gains resolved inequities between these two groups.}

\topic{For equity, these inclusivity gains for the \personaVal{white}{lower} self-efficacy engineers' ratings increased their equity with their \personaVal{white}{higher} self-efficacy peers.}
As Figure~\ref{fig:02_se_equity_changes_same_val} shows, the inclusivity gains for the disadavantaged group occurred without detriment to the advantaged group, achieving nearly perfect equity.

\input{figures/dv_equity_changes_same_val/02_se_equity_changes_same_val}


\subsection{Pattern 3: Decreased inclusivity for either problem-solving style value $\rightarrow$ decreased equity%
\draftStatus{AAA}{2.5+}}
\label{subsec:less_for_one_less_equity}

\boldify{The last pattern was one of the ``worst case'' scenarios for inclusivity and equity---where the adaptation supposed to help one group \textit{decreased} inclusivity, impacting equity}

\topic{This last pattern was one of the ``worst case'' scenarios for inclusivity and equity (Table~\ref{tab:12_inclusive_equity_match_story_truth_table}, rows 4 \& 5).}
It occurred when the LLM's adaptations, which tried to match engineers' problem-solving style values, actually \textit{decreased} inclusivity for either the ``Abi''-like or ``Tim''-like engineers' values (only one `$\downarrow$' in the middle).
This decreased inclusivity decreased equity between the two groups (a `$\downarrow$' in the last column).

\boldify{Figure~\ref{fig:03_ip_mot_inclusion_for_me} shows an example of this pattern for the engineers in the Motivations treatment}

\topic{Figure~\ref{fig:03_mot_inclusion_for_me} shows how the LLM's explanations that matched engineers' motivations could change inclusivity unexpectedly.}
These explanations, which should have increased inclusivity like they had for the other engineers in  Sections~\ref{subsec:more_for_both_more_equity} and~\ref{subsec:more_for_one_more_equity}, actually \textit{decreased} it for both the \personaVal{AbiOrange}{task-oriented} (left) \textit{and} \personaVal{TimBlue}{tech-oriented} (right) engineers.
However, these decreases did not occur evenly across engineer groups;
\llamaWithVal{AbiOrange}{Mot}{Task} decreased \personaVal{white}{task-oriented} engineers' ratings four times, but \llamaWithVal{TimBlue}{Mot}{Tech} decreased the \personaVal{white}{tech-oriented} engineers' ratings \textit{nine} times, occurring almost evenly across \textit{all four} rating categories.


\input{figures/dv_inclusion_for_me_figures/03_mot_inclusion_for_me}

\boldify{Some of the tech-oriented engineers provided some insights to explain why \llamaWithVal{white}{Mot}{Tech} decreased their ratings.}

\topic{Some tech-oriented engineers provided potential insights to explain why their \llamaWithVal{white}{Mot}{Tech} ratings may have decreased.}
Table~\ref{tab:03_llm_mot_adapt_greeting} highlights how the LLM adapted its explanations to greet \personaVal{white}{tech-oriented} engineers, and two engineers independently expressed concerns about the rock-paper-scissors and Fibonacci sequence greeting:

\input{tables/LLM_prob_solve_adaptations/03_llm_mot_adapt_greeting}

\quoteWithLLM{TimBlue}{04}{Mot}{Tech}{TimBlue}{llmSameValue}{It did contain unnecessary details, such as this whole intro of `Hi there, tech enthusiasts' and also just saying like `A COBOL Program!'... I'm not a fan of it.}

\quoteWithLLM{TimBlue}{26}{Mot}{Tech}{TimBlue}{llmSameValue}{I don't like this much [highlights `A COBOL enthusiast, eh?'], because I'm \textit{not} a COBOL enthusiast... If I am, I'm not asking an LLM to explain it to me... }
P18, who saw the Infinite Loop greeting, brought up how the LLM's greeting came across as friendly twice.
The first time came while they were reading the explanation: 
\quoteWithLLM{TimBlue}{18}{Mot}{Tech}{TimBlue}{llmSameValue}{I like this guy so far \textbf{*pause*} Oh, they got me with the psychology, the friendly introduction, now I'm feeling friendly about [the LLM]. They got me.}
P18 acknowledged that the greeting informed their ratings, acknowledging that they suspected being manipulated:
\quoteWithLLM{TimBlue}{18}{Mot}{Tech}{TimBlue}{llmSameValue}{Yeah, I can recognize that it was the slightly more friendly tone that gave me a positive opinion but that doesn't mean it doesn't work. I can see what it's consciously doing to me... doesn't mean it doesn't work.}

\boldify{Although both motivations had decreased inclusivity, it happened so frequently for tech-oriented motivations that it decreased equity.}

\input{figures/dv_equity_figures/05_mot_equity_changes}

\topic{Although both LLM adaptations had decreased inclusivity for these engineers, \llamaWithVal{white}{Mot}{Tech}'s particularly negative impact on inclusivity for \personaVal{white}{tech-oriented} engineers led to a decrease in equity.}
Figure~\ref{fig:05_mot_equity_changes} shows that with the unadapted LLM, the \personaVal{white}{task-oriented} were more inequitably disadvantaged, relative to the \personaVal{white}{tech-oriented}.
However, the decreases in \personaVal{white}{tech-oriented} engineers' ratings shifted the equity landscape;
the \personaVal{white}{tech-oriented} engineers now had \textit{all} of the inequitable disadvantages, relative to the \personaVal{white}{task-oriented}.
Results like these suggest the worst-case scenario, where disparate decreases to inclusivity against one group of problem-solving style values decreases equity.


%% file: tables/12_inclusive_equity_match_story_truth_table.tex
\begin{table}[h]
    \centering
     \includegraphics[width = 0.48\linewidth]{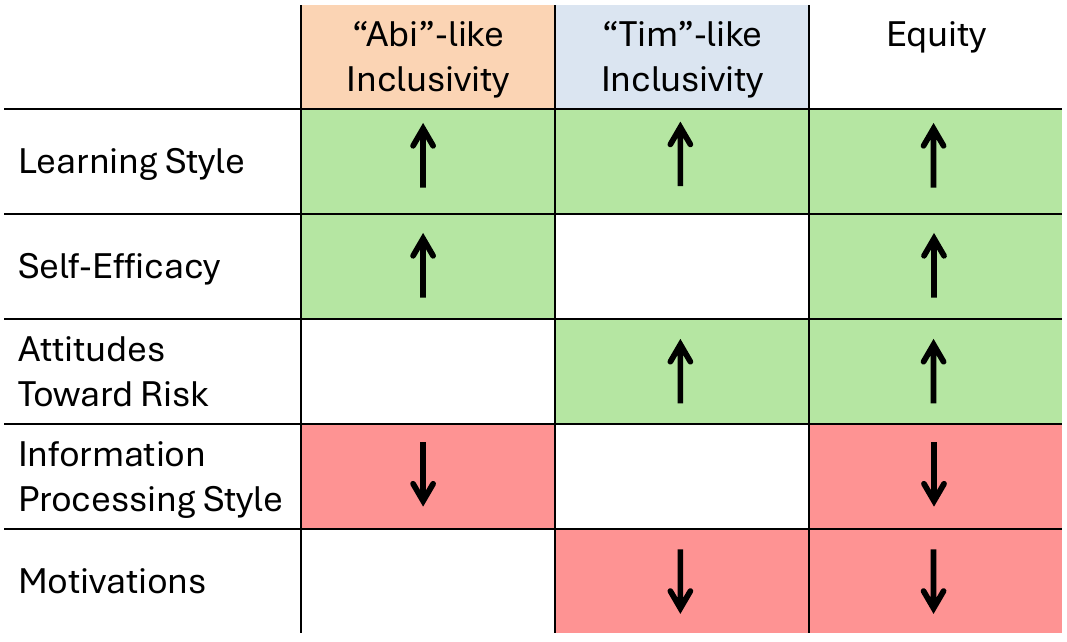}
    \caption{How the LLM's matching adaptations changed inclusivity (middle columns) and equity (right column) for engineers' five problem-solving types (rows).
    Three patterns emerged from engineers' data.
    $\uparrow$/$\downarrow$ : positive/negative outcomes for inclusivity and equity.
    Blank : no change to inclusivity.}
    
    \label{tab:12_inclusive_equity_match_story_truth_table}
\end{table}

%% file: figures/dv_inclusion_for_me_figures/01_learn_inclusion_for_me.tex
\begin{figure}[]
    \centering
    \begin{tabular}{cc}
    Inclusivity for \personaVal{AbiOrange}{process} with \llamaWithVal{AbiOrange}{Learn}{Process} & Inclusivity for \personaVal{TimBlue}{tinker} with \llamaWithVal{TimBlue}{Learn}{Tinker}\\
    \includegraphics[width =.48\linewidth]{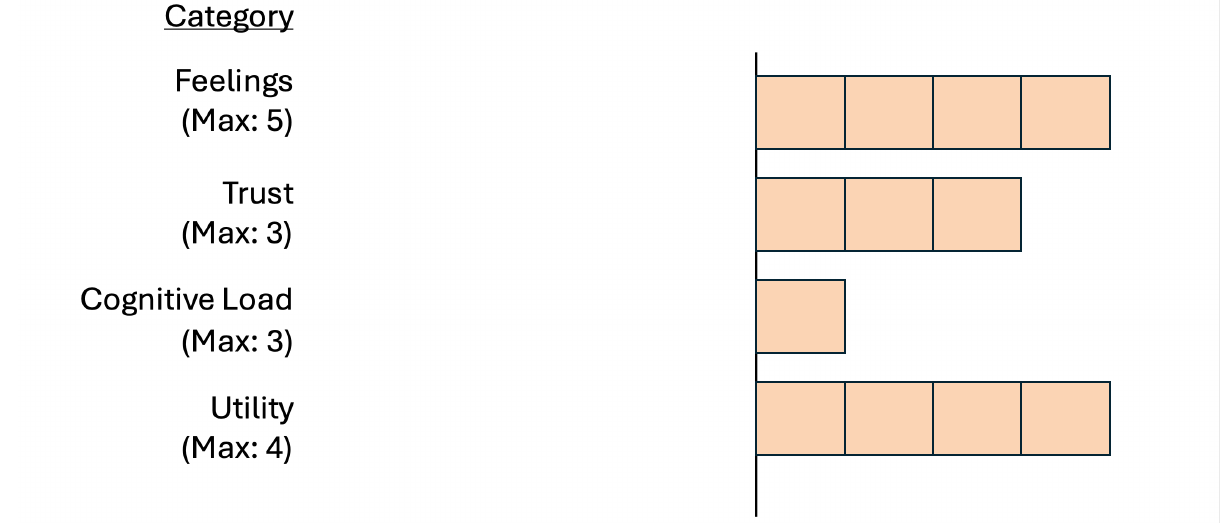} &
     \includegraphics[width=.48\linewidth]{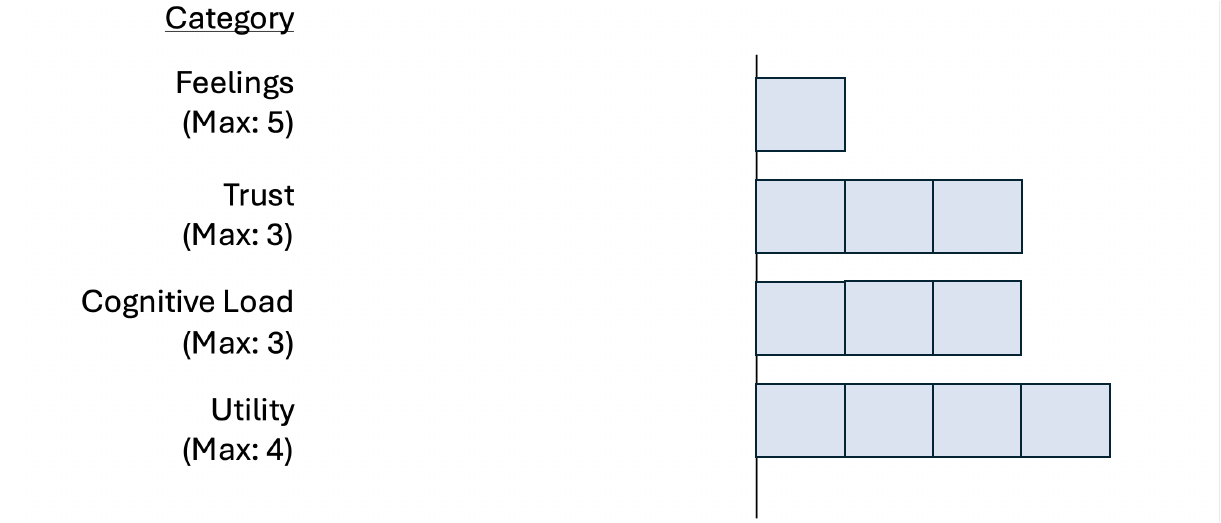}  
     \end{tabular}
    \caption{Inclusivity increases ($\rightarrow$) and decreases ($\leftarrow$) for \personaVal{AbiOrange}{process-oriented} and \personaVal{TimBlue}{tinkering-oriented} learners' ratings.
    \llamaWithVal{AbiOrange}{Learn}{Process} increased inclusivity in 12 of the process-oriented learners' ratings.
    \llamaWithVal{TimBlue}{Learn}{Tinker} increased inclusivity in 11 of the tinkering-oriented learners' ratings.}
    \label{fig:01_learn_inclusion_for_me}
\end{figure}

%% file: tables/LLM_prob_solve_adaptations/01_llm_learning_style_adapt_sections.tex
\begin{table}[]
    \centering
    \begin{tabular}{p{0.47\linewidth}p{0.47\linewidth}}
    \toprule
    \multicolumn{1}{c}{Unadapted LLM} & \multicolumn{1}{c}{\llamaWithVal{AbiOrange}{Learn}{Process}} \\
    \midrule
    \includegraphics[width = \linewidth]{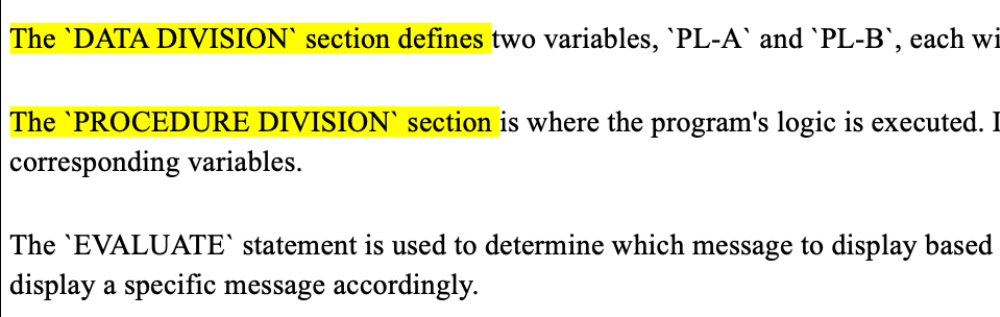} & \includegraphics[width = \linewidth]{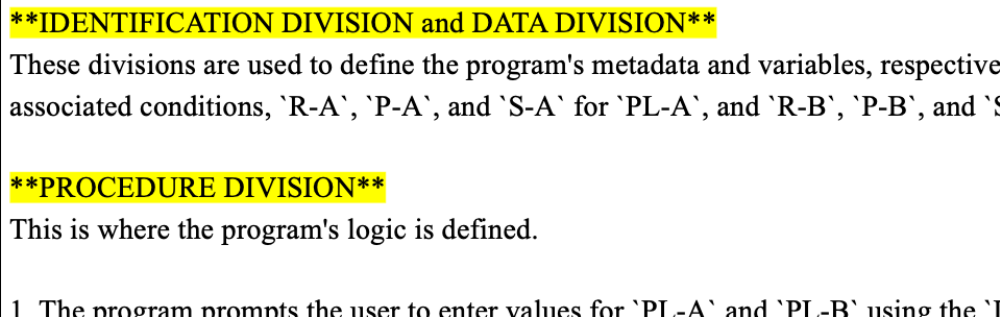}\\
    \bottomrule
    \end{tabular}
    \caption{Screenshots from the unadapted LLM (left) and the adaptated LLM for \personaVal{AbiOrange}{process-oriented} learners (\llamaWithVal{white}{Learn}{Process}, right), explaining the Rock-Paper-Scissors program (highlights added to show the correspondences).
    The LLM adapted to process-oriented learners by explaining the purpose of the different sections, which was helpful to process-oriented learners like \personaVal{AbiOrange}{P31} (see text).
    }
    \label{tab:01_llm_learning_style_adapt_sections}
\end{table}

%% file: figures/dv_equity_changes_same_val/01_learn_equity_changes_same_val.tex
\begin{figure}[]
    \centering
    \begin{tabular}{cc}
    Unadapted LLM Inequity & Adapted LLM (same value) Inequity \\
    \includegraphics[width =.48\linewidth]{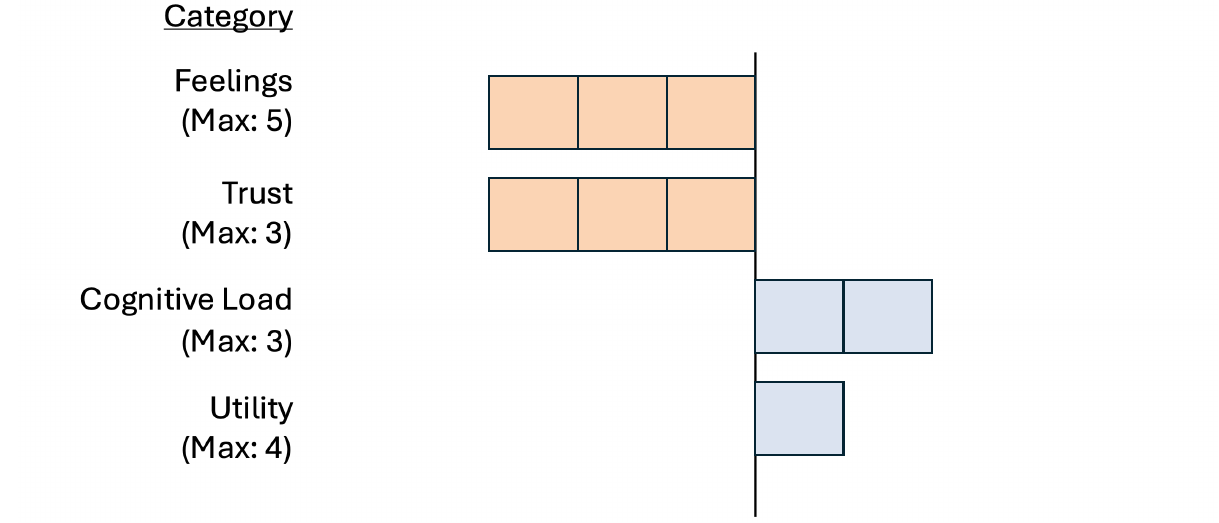} &
     \includegraphics[width=.48\linewidth]{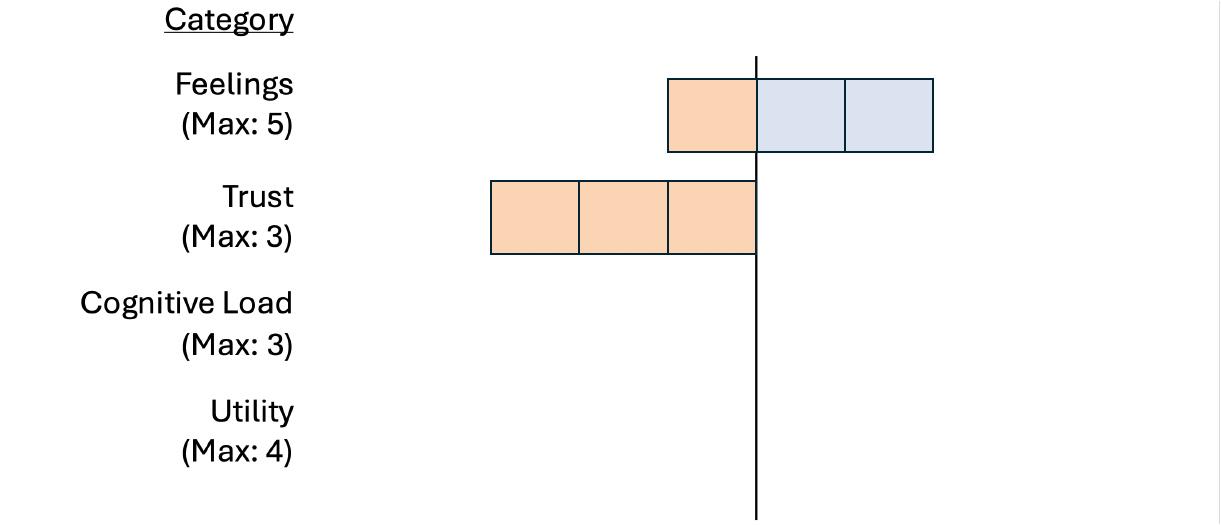}  
     \end{tabular}
    \caption{Rating inequities (boxes) between \personaVal{AbiOrange}{process-oriented} and \personaVal{TimBlue}{tinkering-oriented} learners for the unadapted LLM (left) and the same-value learning-style adapted LLMs (right).
    A majority of the unadapted LLM disadvantages went against the process-oriented learners.
    However, the inclusivity benefits (Figure~\ref{fig:01_learn_inclusion_for_me}) improved equity, eliminating inequities in Cognitive Load and Utility.}
    \label{fig:01_learn_equity_changes_same_val}
\end{figure}

%% file: figures/dv_inclusion_for_me_figures/02_se_inclusion_for_me.tex
\begin{figure}[h]
    \centering
    \begin{tabular}{cc}
   Inclusivity for \personaVal{AbiOrange}{lower} with \llamaWithVal{AbiOrange}{SelfEff}{Lower} & Inclusivity for \personaVal{TimBlue}{higher} with \llamaWithVal{TimBlue}{SelfEff}{Higher}\\
     \includegraphics[width =.48\linewidth]{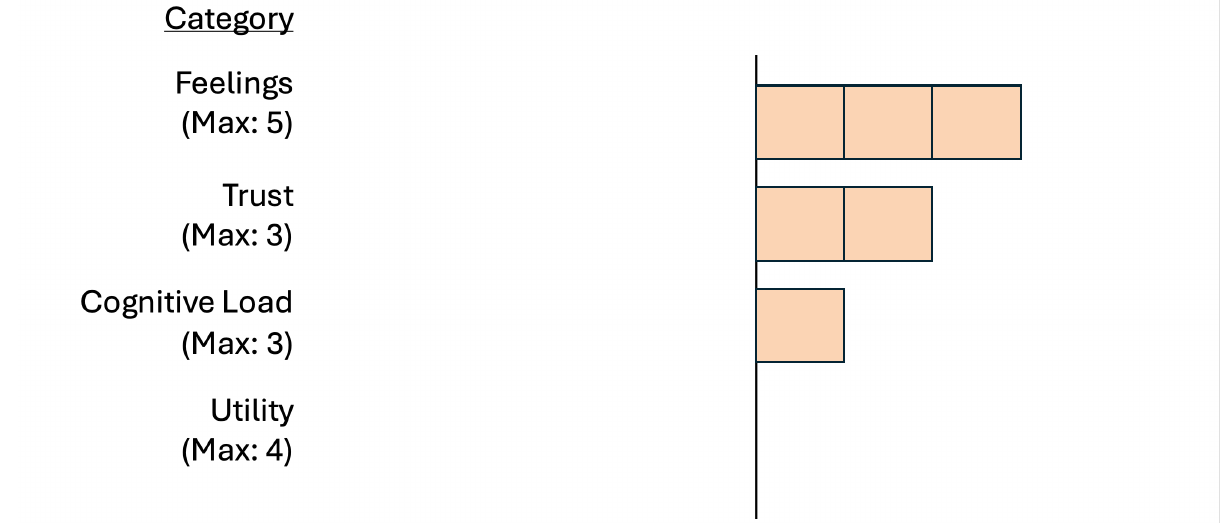} &
     \includegraphics[width=.48\linewidth]{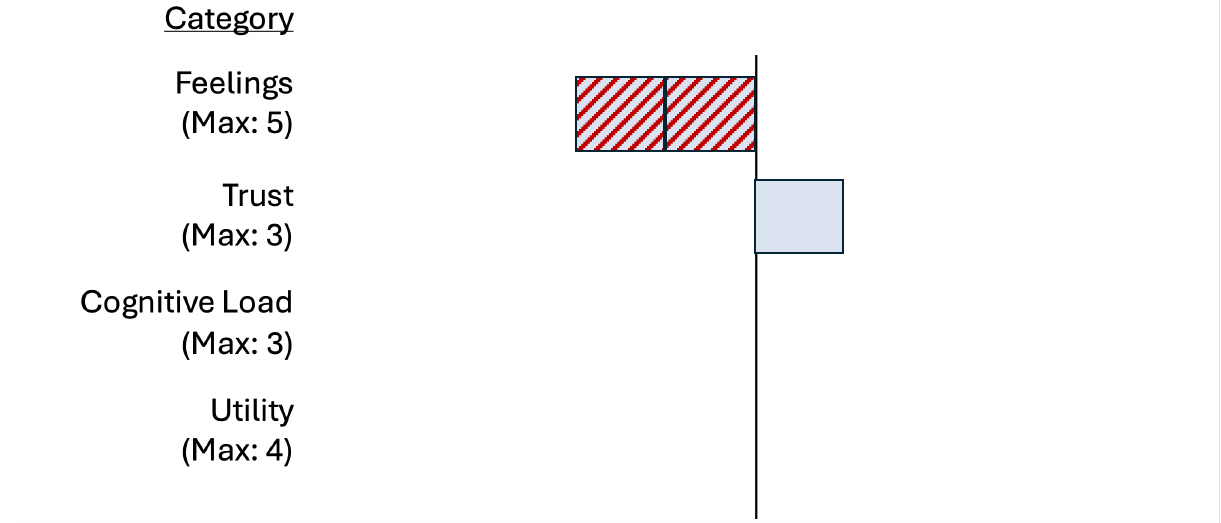}   
     
    \end{tabular}
    \caption{The number of inclusivity increases and decreases for engineers with \personaVal{AbiOrange}{lower} (left) and \personaVal{TimBlue}{higher} (right) computer self-efficacy.
    \llamaWithVal{AbiOrange}{SelfEff}{Lower} increased inclusivity more frequently for the  \personaVal{white}{lower} self-efficacy engineers than \llamaWithVal{TimBlue}{SelfEff}{Higher} did for their \personaVal{white}{higher} self-efficacy peers.
    }
    \label{fig:02_se_inclusion_for_me}
\end{figure}

%% file: tables/LLM_prob_solve_adaptations/02_llm_se_adapt_supportive_language.tex
\begin{table}[b]
    \centering
    \begin{tabular}{c}
    \toprule
    Unadapted LLM\\ 
    \midrule
         \includegraphics[width = \linewidth]{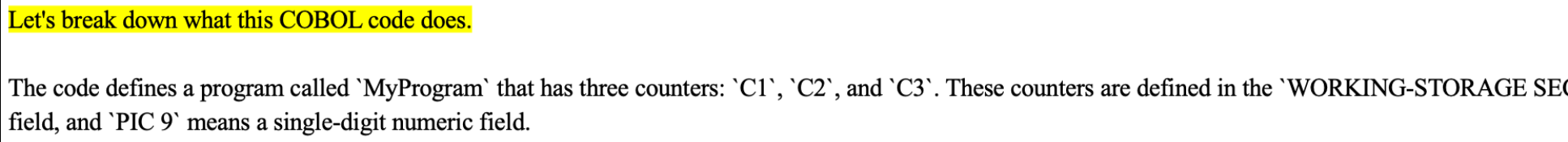}  \\
    \toprule
    \llamaWithVal{AbiOrange}{SelfEff}{Lower}\\ 
    \midrule
         \includegraphics[width = \linewidth]{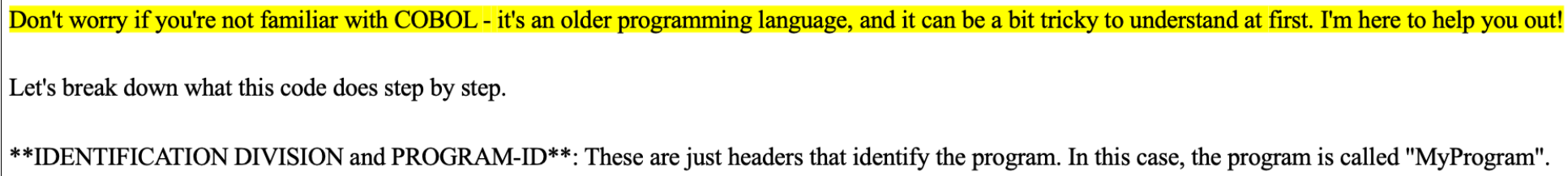}  \\ 
    \toprule
    \llamaWithVal{TimBlue}{SelfEff}{Higher}\\
    \midrule
    \includegraphics[width = \linewidth]{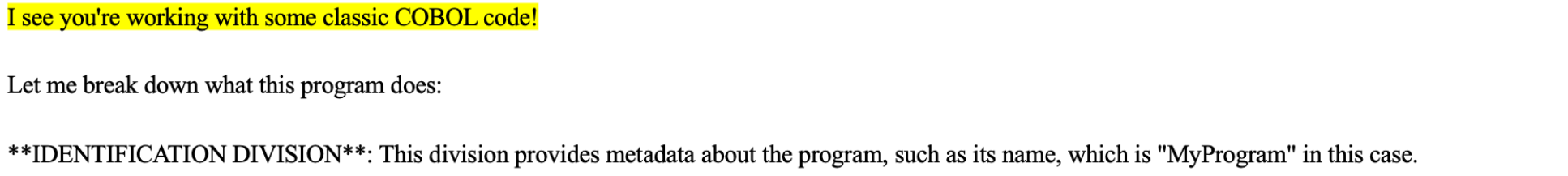}  \\ 
    \bottomrule
    \end{tabular}
    \caption{Portions of the explanations from the unadapted LLM (top), the adapted LLM for \personaVal{AbiOrange}{lower} self-efficacy engineers (\llamaWithVal{white}{SelfEff}{Lower}, middle), and the adapted LLM for \personaVal{TimBlue}{higher} self-efficacy engineers (\llamaWithVal{white}{SelfEff}{Higher}, bottom), explaining the Infinite Loop program.
    Highlights added to show the differences.
    Although the LLM's adaptation to \personaVal{TimBlue}{higher} self-efficacy engineers was minimal, the LLM adapted to \personaVal{AbiOrange}{Lower} self-efficacy engineers by using more supportive language.}
    \label{tab:02_llm_se_adapt_supportive_language}
\end{table}

%% file: figures/dv_equity_changes_same_val/02_se_equity_changes_same_val.tex
\begin{figure}[]
    \centering
    \begin{tabular}{cc}
    Unadapted LLM Inequity & Adapted LLM (same value) Inequity \\
     \includegraphics[width =.48\linewidth]{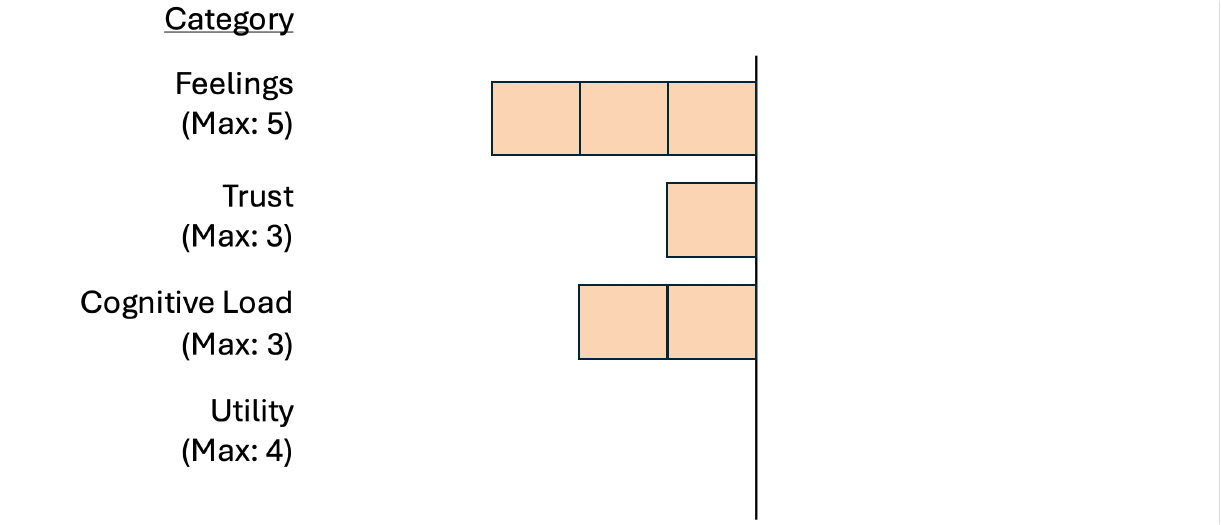} &
     \includegraphics[width=.48\linewidth]{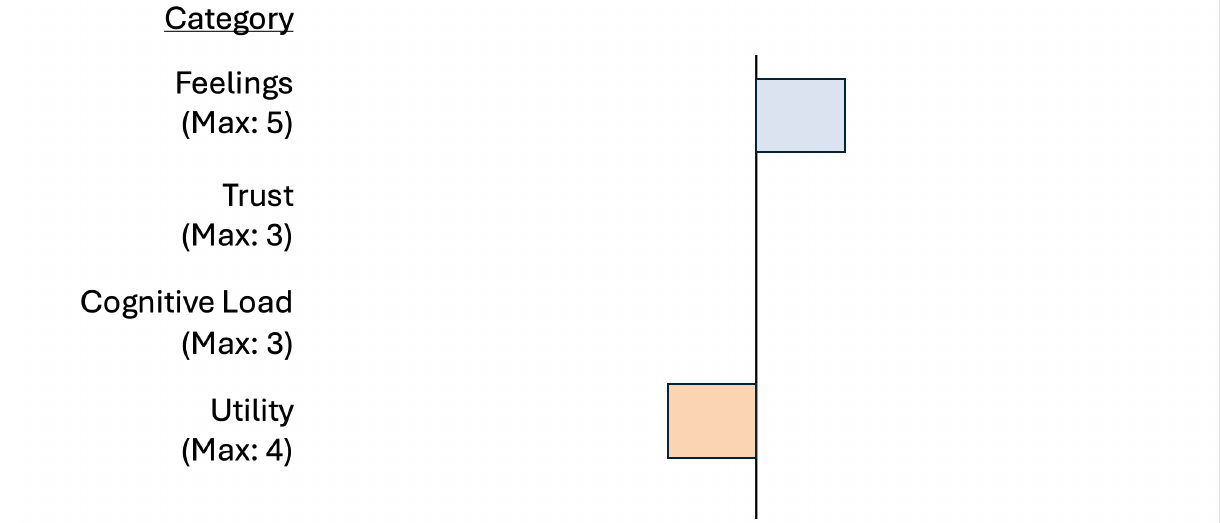}   
    \end{tabular}
    \caption{The number of inequities for the unadapted LLM (left) and the self-efficacy adapted LLMs (right) between engineers with \personaVal{AbiOrange}{lower} and \personaVal{TimBlue}{higher} self-efficacy values.
    For the unadapted LLM, \textit{all} of the disadvantages went against the \personaVal{white}{lower} self-efficacy engineers.
    However, \llamaWithVal{AbiOrange}{SelfEff}{Lower} increased inclusivity so frequently (Figure~\ref{fig:02_se_inclusion_for_me}) that inequities were almost erased.}
    \label{fig:02_se_equity_changes_same_val}
\end{figure}

%% file: figures/dv_inclusion_for_me_figures/03_mot_inclusion_for_me.tex
\begin{figure}[b]
    \centering
    \begin{tabular}{cc}
    Inclusivity for \personaVal{AbiOrange}{task} with \llamaWithVal{AbiOrange}{Mot}{Task} & Inclusivity for \personaVal{TimBlue}{tech} with \llamaWithVal{TimBlue}{Mot}{Tech}\\
     \includegraphics[width = .48\linewidth]{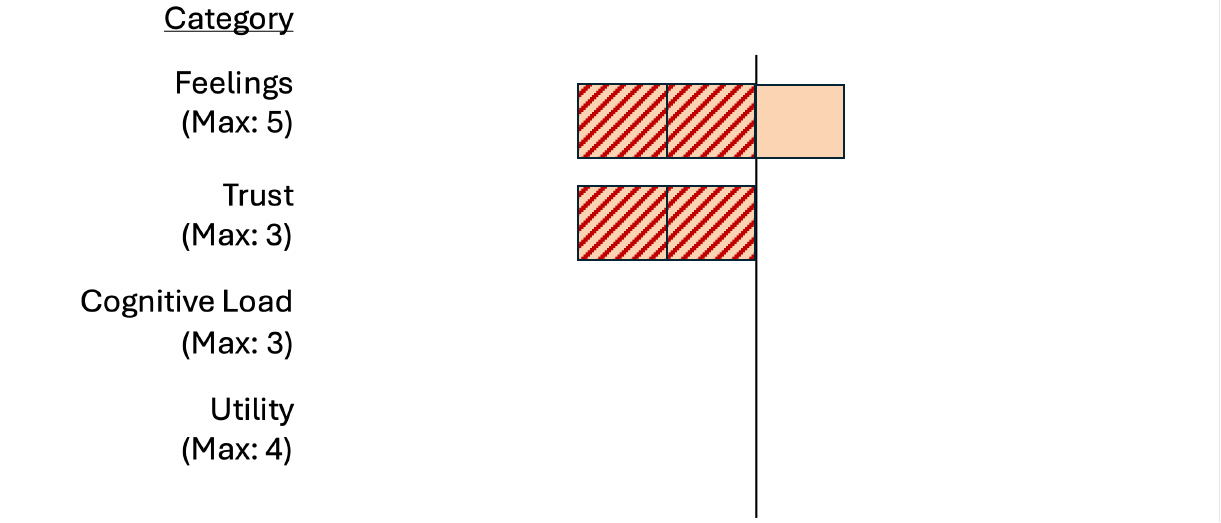}&
     \includegraphics[width=.48\linewidth]{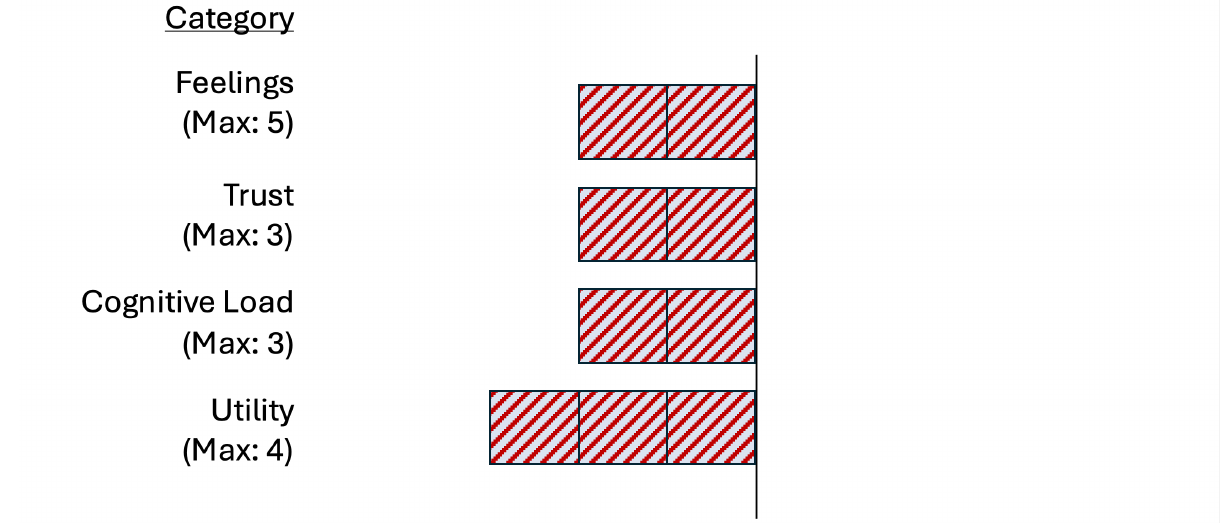}
    \end{tabular}
    \caption{The number of inclusivity increases/decreases for engineers with \personaVal{AbiOrange}{task-oriented} (left) and \personaVal{TimBlue}{tech-oriented} (right) motivations.
    In this treatment, both \llamaWithVal{AbiOrange}{Mot}{Task} and \llamaWithVal{TimBlue}{Mot}{Tech} unexpectedly \textit{decreased} inclusivity more than they increased it for the engineers they adapted for.
    However, \llamaWithVal{white}{Mot}{Tech} decreased inclusivity more frequently.}
    \label{fig:03_mot_inclusion_for_me}
\end{figure}

%% file: tables/LLM_prob_solve_adaptations/03_llm_mot_adapt_greeting.tex
\begin{table}[h]
    \centering
    \begin{tabular}{c}
    \toprule
    Rock-Paper-Scissors\\ 
    \midrule
         \includegraphics[width = 0.55\linewidth]{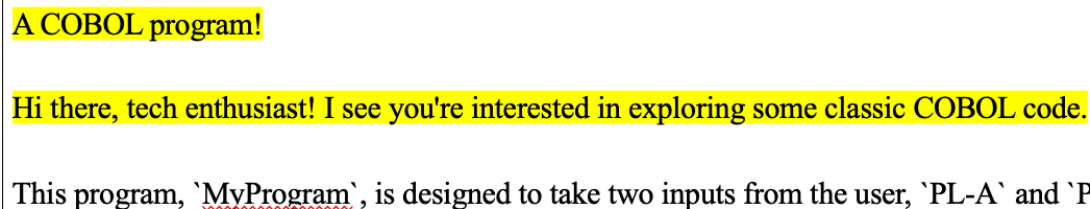}  \\
    \toprule
    Fibonacci Sequence\\ 
    \midrule
         \includegraphics[width = 0.55\linewidth]{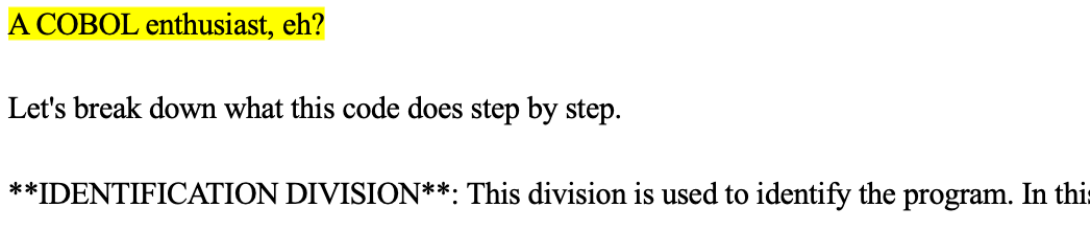}  \\ 
    \toprule
    Infinite Loop\\
    \midrule
    \includegraphics[width = 0.556\linewidth]{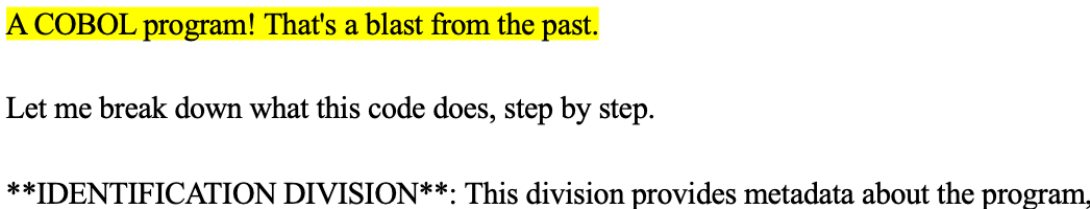}  \\ 
    \bottomrule
    \end{tabular}
    \caption{Portions of how the adapted LLM for engineers with \personaVal{TimBlue}{tech-oriented} motivations explained the rock-paper-scissors (top), Fibonacci sequence (middle), and Infinite Loop (bottom), explaining the Infinite Loop program.
    Highlights added to show the differences.
    \personaVal{TimBlue}{P04}, \personaVal{TimBlue}{P26}, and \personaVal{TimBlue}{P18} had mixed reactions to these greetings.}
    \label{tab:03_llm_mot_adapt_greeting}
\end{table}

%% file: figures/dv_equity_figures/05_mot_equity_changes.tex
\begin{figure}[b]
    \centering
    \begin{tabular}{cc}
    Unadapted LLM Inequity & Adapted LLM (same value) Inequity \\
     \includegraphics[width =.48\linewidth]{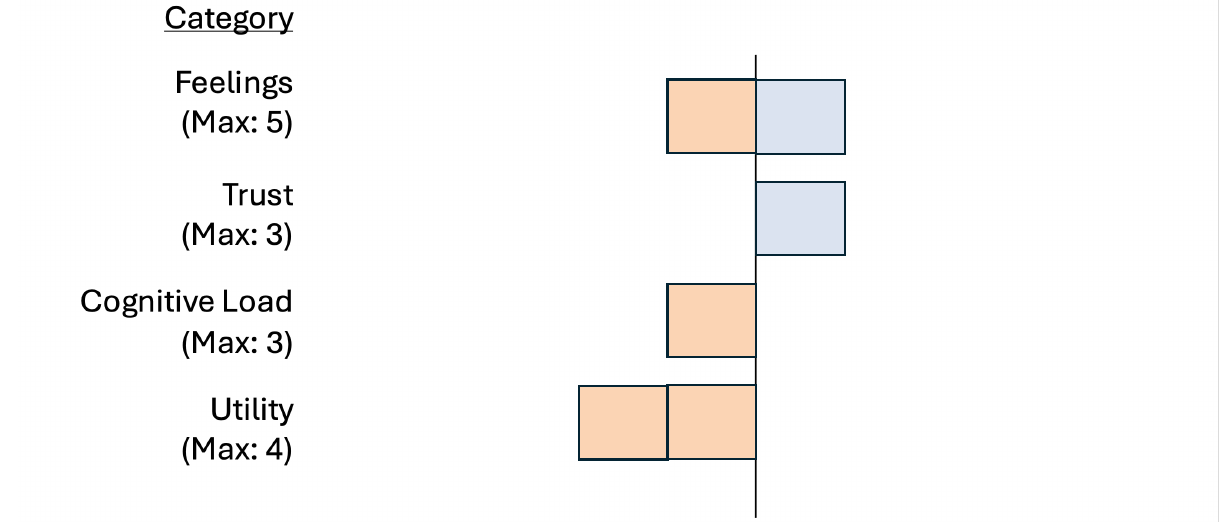} &
     \includegraphics[width=.48\linewidth]{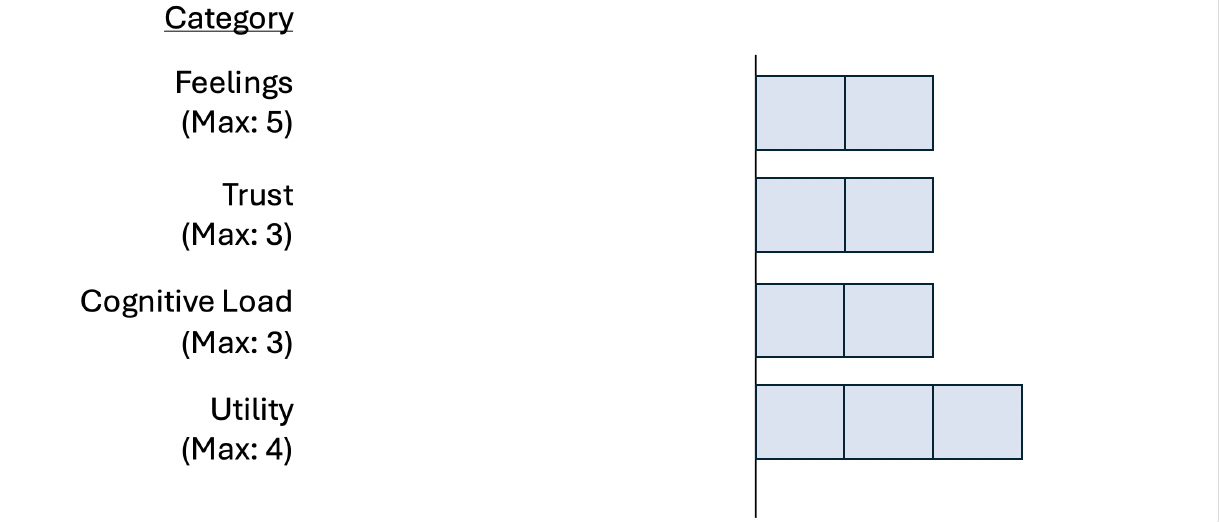}   
    \end{tabular}
    \caption{The number of inequities for the unadapted LLM (left) and the motivations adapted LLMs (right) between engineers with \personaVal{AbiOrange}{task-oriented} and \personaVal{TimBlue}{tech-oriented} motivations.
    The unadapted LLM disadvantaged the \personaVal{white}{task-oriented} engineers more frequently.
    However, \llamaWithVal{TimBlue}{Mot}{Tech} had \textit{decreased} inclusivity so frequently (Figure~\ref{fig:03_mot_inclusion_for_me}) that the \personaVal{white}{\textit{tech}-oriented} engineers were disadvantaged.}
    \label{fig:05_mot_equity_changes}
\end{figure}

%% file: doc/04_2_results_rq_equity_inclusion_mismatching_vals.tex
\section{Results: The LLM's Adaptation Mismatches---Were They \textit{Un}inclusive \& \textit{In}equitable?
\draftStatus{(TOP)---AAA}{2.8}}
\label{sec:adapt_mismatch}

\input{tables/13_inclusive_equity_mismatch_story_truth_table}

\topic{\textbf{RQ3} asked how an LLM's adaptations that \textit{mis}matched diverse engineers' problem-solving style values changed inclusivity and equity, relative to unadapted explanations.}
Table~\ref{tab:13_inclusive_equity_mismatch_story_truth_table} provides the bottom line for ``Abi''-like and ``Tim''-like engineers.
Engineers' ratings showed two \textit{new} patterns in these data across the five problem-solving style types:

\begin{enumerate}[leftmargin = 0.5cm, label = (\arabic*)]
    \item Increasing inclusivity for one problem-solving style value decreased equity (Learning Style).
    \item Decreasing inclusivity for both problem-solving style values decreased equity (Self-Efficacy and Motivations).
\end{enumerate}
The same three problem-solving style types featured in these two patterns again, so these subsections discuss both patterns for mismatching explanations, drawing comparisons between the inclusivity and equity findings from Section~\ref{sec:adapt_match}.

\subsection{Pattern 1: Increased inclusivity for one problem-solving style value $\rightarrow$ decreased equity
\draftStatus{AAA}{2.5}}

\boldify{This first pattern highlighted a potential drawback for explanations which adapted for the opposite engineers' problem-solving style value}

\topic{This first pattern highlighted a potential drawbacks for inclusivity when explanations adapted for the \textit{opposite} engineers' problem-solving style values (Table~\ref{tab:13_inclusive_equity_mismatch_story_truth_table}, first row).}
In this pattern, these explanations still increased inclusivity for the ``Tim''-like engineers (one `$\uparrow$' in the middle columns).
However, this increased inclusivity led to \textit{decreased} equity between the two groups (a `$\downarrow$' in the last column).

\boldify{Figure~\ref{fig:01_learn_inclusion_not_for_me} shows how the opposite adapting explanations for diverse learning styles led to disparate outcomes for inclusivity for process-oriented learners (Section~\ref{subsec:more_for_both_more_equity}).}

\topic{Figure~\ref{fig:01_learn_inclusion_not_for_me} highlights this pattern, showing how mismatching explanations changed inclusivity for engineers' learning styles.}
Contrary to the findings in Section~\ref{subsec:more_for_both_more_equity}, the left of the figure shows that for  \personaVal{AbiOrange}{process-oriented} learners, \llamaWithVal{TimBlue}{Learn}{Tinker} did not increase inclusivity as frequently as \llamaWithVal{AbiOrange}{Learn}{Process} had;
\llamaWithVal{white}{Learn}{Tinker} increased inclusivity for engineers' ratings only four times across the Feelings and Utility categories, eight fewer than \llamaWithVal{white}{Learn}{Process} had provided them.

\input{figures/dv_inclusion_not_for_me_figures/01_learn_inclusion_not_for_me}

\boldify{Some of process-oriented learners' quotes even pointed out that they thought it was inferior, helping to explain why they might have felt that th mismatching explanations were less supportive.}

\topic{Some of the process-oriented learners' comments reflected what they found less helpful about \llamaWithVal{white}{Learn}{Tinker}.}
P31 compared their feeling of certainty between \llamaWithVal{white}{Learn}{Process} and \llamaWithVal{white}{Learn}{Tinker}:

\quoteWithLLM{AbiOrange}{31}{Learn}{Process}{TimBlue}{llmOppositeValue}{I do feel a lot more uncertain than I did before [with \llamaWithVal{AbiOrange}{Learn}{Process}]}
and stated that they did not like that the LLM adaptation had not been as organized: 
\quoteWithLLM{AbiOrange}{31}{Learn}{Process}{TimBlue}{llmOppositeValue}{I didn't like its organization as much as [\llamaWithVal{AbiOrange}{Learn}{Process}]. Maybe I'm rating [\llamaWithVal{TimBlue}{Learn}{Tinker}] a little bit harshly, but I think it was inferior...}

Other process-oriented participants, who had praised the \llamaWithVal{white}{Learn}{Process}, used their prior experience with that LLM to criticize \llamaWithVal{white}{Learn}{Tinker}.
In Section~\ref{subsec:more_for_both_more_equity}, P22 praised how step-by-step \llamaWithVal{white}{Learn}{Process}'s explanation was.
However, P22 stated how \llamaWithVal{white}{Learn}{Tinker}'s explanation dove in too deep:
\quoteWithLLM{AbiOrange}{22}{Learn}{Process}{TimBlue}{llmOppositeValue}{I'm feeling like I'm now too deep into the counting and the logic and whatever and do not know what the loop procedure does.}
P22 later resurrected their concerns about how \llamaWithVal{white}{Learn}{Tinker} had explained the loop structure to express that the explanation had unnecessary details:
\quoteWithLLM{AbiOrange}{22}{Learn}{Process}{TimBlue}{llmOppositeValue}{This is completely wrong. It was unnecessary to go [that way] about the [loop] iterations.}

\boldify{The right side of Figure~\ref{fig:01_learn_inclusion_not_for_me} shows that \llamaWithVal{AbiOrange}{Learn}{Process} still increased inclusivity for \personaVal{TimBlue}{tinkering-oriented} learners' ratings}

\topic{In contrast, the right side of Figure~\ref{fig:01_learn_inclusion_not_for_me} shows that \llamaWithVal{AbiOrange}{Learn}{Process} still increased inclusivity for \personaVal{TimBlue}{tinkering-oriented} learners ratings.}
Although the LLM adapted to the \textit{opposite} learning style, \llamaWithVal{white}{Learn}{Process} was almost as effective as \llamaWithVal{white}{Learn}{Tinker} for \personaVal{white}{tinkering-oriented} learners, increasing inclusivity in 10 of their ratings.
This was unexpected, since \llamaWithVal{white}{Learn}{Process} provided only \textit{one} fewer inclusive rating increase than \llamaWithVal{white}{Learn}{Tinker} provided them  (Section~\ref{subsec:more_for_both_more_equity}).

\boldify{These differences in inclusivity \textit{also} led to changes in equity. }

These differences in inclusivity, compared to Section~\ref{subsec:more_for_both_more_equity}, led to different outcomes for equity too.
For example, Figure~\ref{fig:01_learn_equity_mismatch_comparison} shows that for process-oriented and tinkering-oriented learners, equity between these groups for the unadapted LLM (left) and the LLM that adapted to their opposite value (right) decreased.
The increases in inclusivity (Figure~\ref{fig:01_learn_inclusion_not_for_me}) provided process-oriented learners with only one fewer inequity. 
However, tinkering-oriented learners' inclusivity gains eliminated their inequities.
Consequently, the only group left with \textit{any} inequity disadvantages left between the two groups were the process-oriented learners, thus reducing overall equity.

\input{figures/dv_equity_changes_opposite_val/01_learn_equity_changes_opposite_val}



\subsection{Pattern 2: Decreased inclusivity for both problem-solving styles $\rightarrow$ decreased equity
\draftStatus{AAA}{0.5}}
\label{subsec:mismatch_6_2}

\boldify{The second pattern in Table~\ref{tab:13_inclusive_equity_mismatch_story_truth_table} involved decreased inclusivity for \textit{both} problem-solving style values and decreasd equity }

The second pattern from Table~\ref{tab:13_inclusive_equity_mismatch_story_truth_table} was a ``worst case'' scenario for inclusive design (Self-Efficacy and Motivations rows).
In this pattern, the explanations that adapted for opposing problem-solving style values decreased inclusivity for both ``Abi''-like and ``Tim''-like engineers (a `$\downarrow$' in both middle columns).
These decreases in inclusivity also led to decreased equity between the two groups (a $\downarrow$ in the last column).

\boldify{Figure~\ref{fig:04_se_inclusion_not_for_me} highlights one instance of this pattern. Unlike Section~\ref{subsec:more_for_one_more_equity}, lower self-efficacy got dunked on.}

One instance of this pattern arose for the engineers in the self-efficacy treatment (Figure~\ref{fig:04_se_inclusion_not_for_me}).
The left half of the figure shows that for the \personaVal{AbiOrange}{lower} self-efficacy engineers, \llamaWithVal{TimBlue}{SelfEff}{Higher} increased their ratings' inclusivity only \textit{once} in the Trust category (five fewer inclusivity increases than \llamaWithVal{white}{SelfEff}{Lower} had provided).
In contrast to the findings from Section~\ref{subsec:more_for_one_more_equity}, where \llamaWithVal{white}{SelfEff}{Lower} decreased their rating inclusivity only once, \llamaWithVal{white}{SelfEff}{Higher} decreased inclusivity for 11 of these engineers' ratings.

\boldify{However, the right side of the figure shows that higher self-efficacy were both helped and hurt by the adaptation which matched the opposite problem-solving style value}

The right half of the figure shows that for \personaVal{TimBlue}{higher} self-efficacy engineers, \llamaWithVal{AbiOrange}{SelfEff}{Lower}  simultaneously provided more increases \textit{and more decreases} than \llamaWithVal{TimBlue}{SelfEff}{Higher} had;
\llamaWithVal{white}{SelfEff}{Lower} increased inclusivity for five of these engineers' ratings across the Feelings, Trust, Utility, and Quality categories, four more increases than they received from \llamaWithVal{white}{SelfEff}{Higher}. 
However, it also decreased inclusivity for six of these engineers' ratings across all four categories, reflecting a worsening outcome with four additional decreases than they received from the LLM adapted to match their problem-solving style value.

\input{figures/dv_inclusion_not_for_me_figures/04_se_inclusion_not_for_me}

\boldify{These drops in inclusivity might look like they impact both, but holy smokes was it bad for lower self-efficacy engineers in terms of equity.}

The differences in inclusivity between when the LLM adaptation matched engineers' self-efficacies (Section~\ref{subsec:more_for_one_more_equity}) and when they did not changed equity, but in the opposite way.
Figure~\ref{fig:02_se_equity_changes_opposite_val} (left) recalls that with the unadapted LLM, the ratings of the engineers with lower self-efficacies were already inequitably disadvantaged, relative to their higher self-efficacy peers;
they were the \textit{only} engineers disadvantaged.
The right side of Figure~\ref{fig:02_se_equity_changes_opposite_val} suggests that providing engineers with adapted LLMs can be bad for all, but this instance of the second pattern reflected a ``worst case'' because decreases in inclusivity led to worsening equity for those who were disparately impacted initially---in this case, the engineers with lower self-efficacy.

\input{figures/dv_equity_changes_opposite_val/02_se_equity_changes_opposite_val}

%% file: tables/13_inclusive_equity_mismatch_story_truth_table.tex
\begin{table}[h]
    \centering
     \includegraphics[width = 0.48\linewidth]{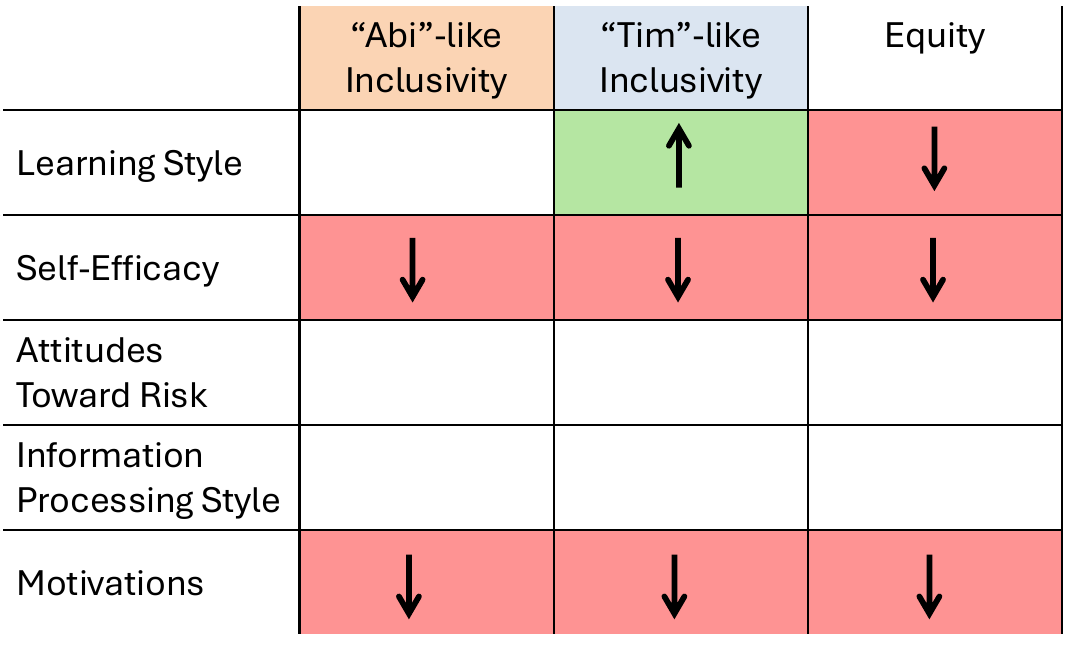}
    \caption{How the LLM's mismatching adaptations changed inclusivity (middle columns) and equity (right column) for engineers' five problem-solving types (rows).
    Two new patterns emerged from engineers' data.
    $\uparrow$/$\downarrow$ : positive/negative outcomes for inclusivity and equity.
    Blank : no change to inclusivity/equity.}
    \label{tab:13_inclusive_equity_mismatch_story_truth_table}
\end{table}

%% file: figures/dv_inclusion_not_for_me_figures/01_learn_inclusion_not_for_me.tex
\begin{figure}[]
    \centering
    \begin{tabular}{cc}
    Inclusivity for \personaVal{AbiOrange}{process} with \llamaWithVal{TimBlue}{Learn}{Tinker} & Inclusivity for \personaVal{TimBlue}{tinker} with \llamaWithVal{AbiOrange}{Learn}{Process}\\
     \includegraphics[width =.48\linewidth]{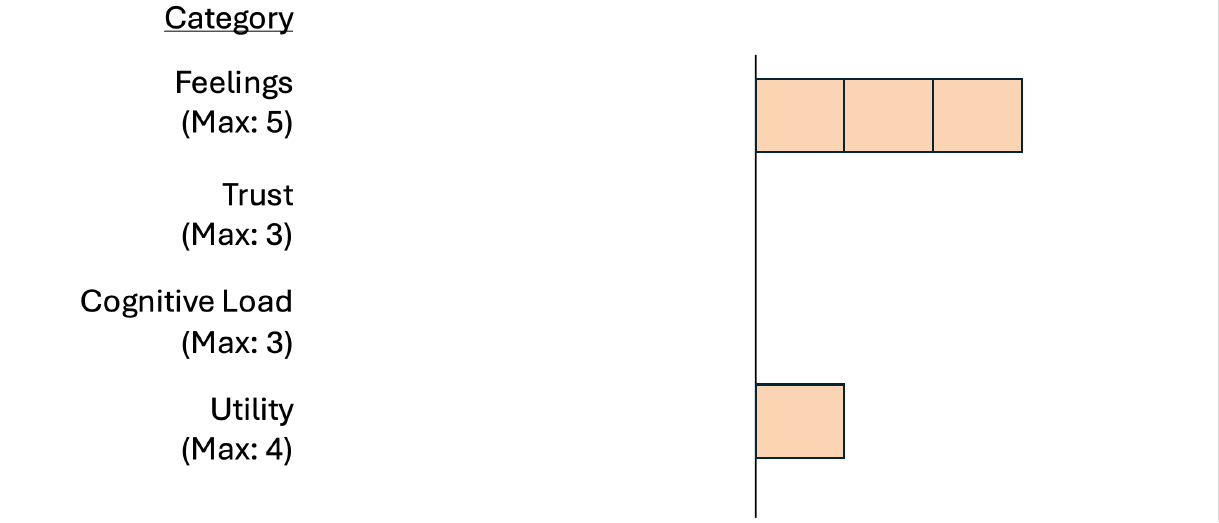} &
     \includegraphics[width=.48\linewidth]{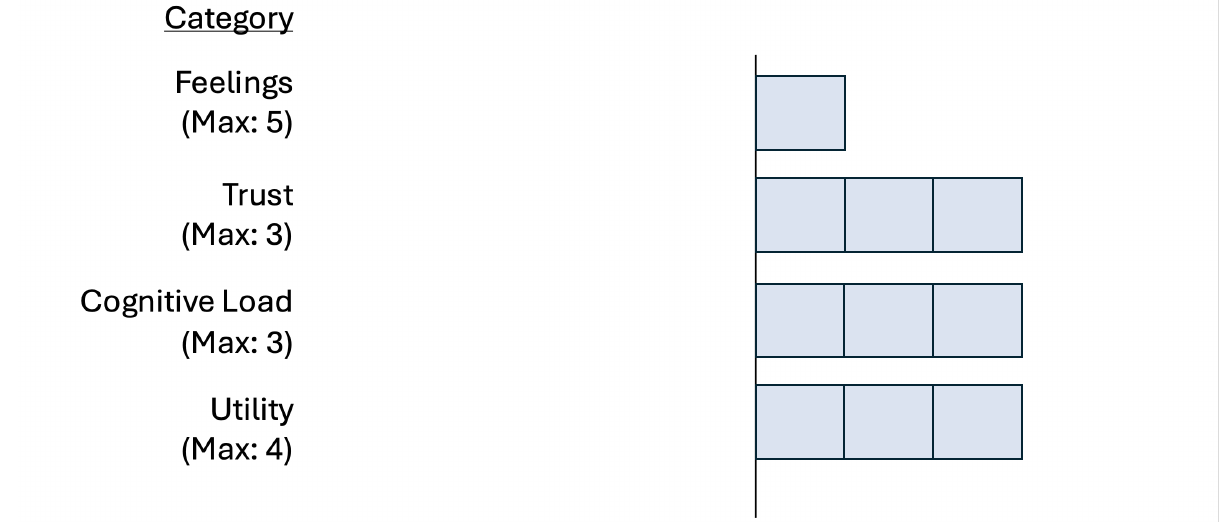}   
    \end{tabular}
    \caption{The number of inclusivity increases/decreases for \personaVal{AbiOrange}{process-oriented} (left) and \personaVal{TimBlue}{tinkering-oriented} (right) learners when the adaptations did not match their learning style.
    Unlike Section~\ref{subsec:more_for_both_more_equity}, \llamaWithVal{TimBlue}{Learn}{Tinker} did not increase inclusivity for the process-oriented engineers' ratings than before.
    However, the \personaVal{TimBlue}{tinkering-oriented} learners seemed unaffected by \llamaWithVal{AbiOrange}{Learn}{Process}.}
    \label{fig:01_learn_inclusion_not_for_me}
\end{figure}

%% file: figures/dv_equity_changes_opposite_val/01_learn_equity_changes_opposite_val.tex
\begin{figure}[]
    \centering
    \begin{tabular}{cp{0.01\linewidth}c}
    Unadapted LLM Inequity & & Adapted LLM (opposite value) Inequity \\
     \includegraphics[width =.47\linewidth]{assets/dv_equity_assets/unadapted_equity_assets/learn_unadapted_equity.pdf} &
     &
     \includegraphics[width=.47\linewidth]{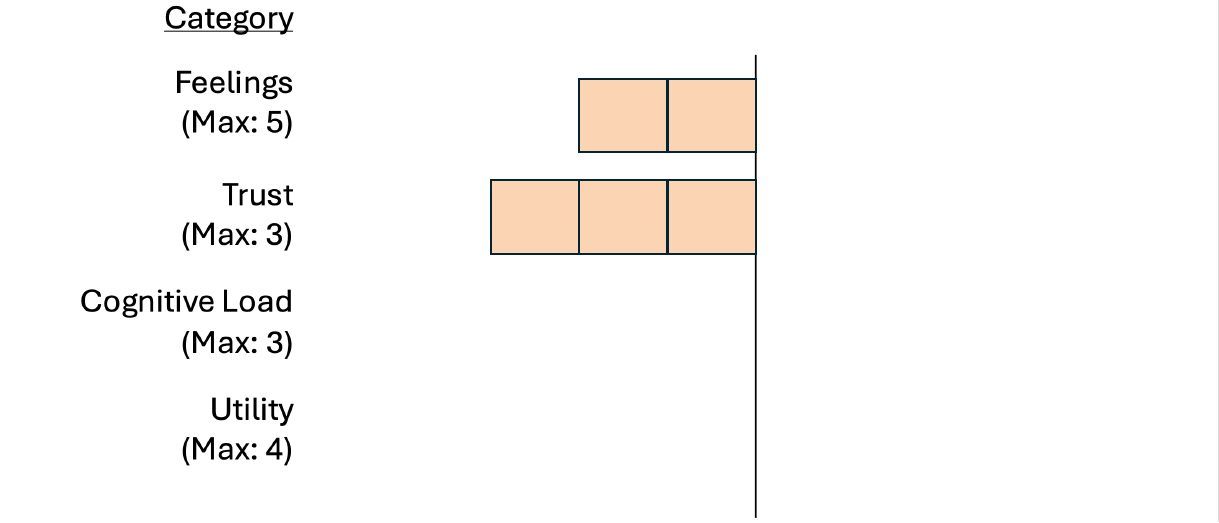}   
    \end{tabular}
    \caption{Rating inequities (boxes) between \personaVal{AbiOrange}{process-oriented} and \personaVal{TimBlue}{tinkering-oriented} learners for the unadapted LLM (left) and the learning-style adapted LLMs which matched the \textit{opposite} value (right).
    Unlike Section~\ref{subsec:more_for_both_more_equity}, the infrequent inclusivity increases for process-oriented learners with \llamaWithVal{white}{Learn}{Tinker} left them similarly disadvantaged as they had been with the unadapted LLM.
    However, the inclusivity increases that \llamaWithVal{white}{Learn}{Process} gave the tinkering-oriented learners eliminated their disadvantages.}
    \label{fig:01_learn_equity_mismatch_comparison}
\end{figure}

%% file: figures/dv_inclusion_not_for_me_figures/04_se_inclusion_not_for_me.tex
\begin{figure}[h]
    \centering
    \begin{tabular}{cc}
    Inclusivity for \personaVal{AbiOrange}{lower} with \llamaWithVal{TimBlue}{SelfEff}{Higher} & Inclusivity for \personaVal{TimBlue}{higher} with \llamaWithVal{AbiOrange}{SelfEff}{Lower}\\
     \includegraphics[width =.48\linewidth]{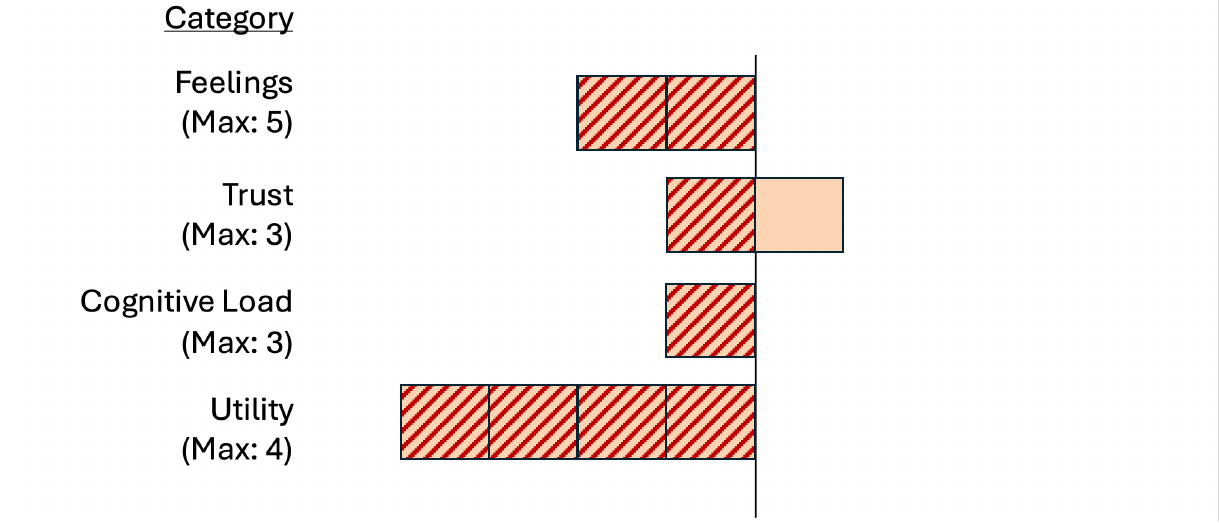} &
     \includegraphics[width=.48\linewidth]{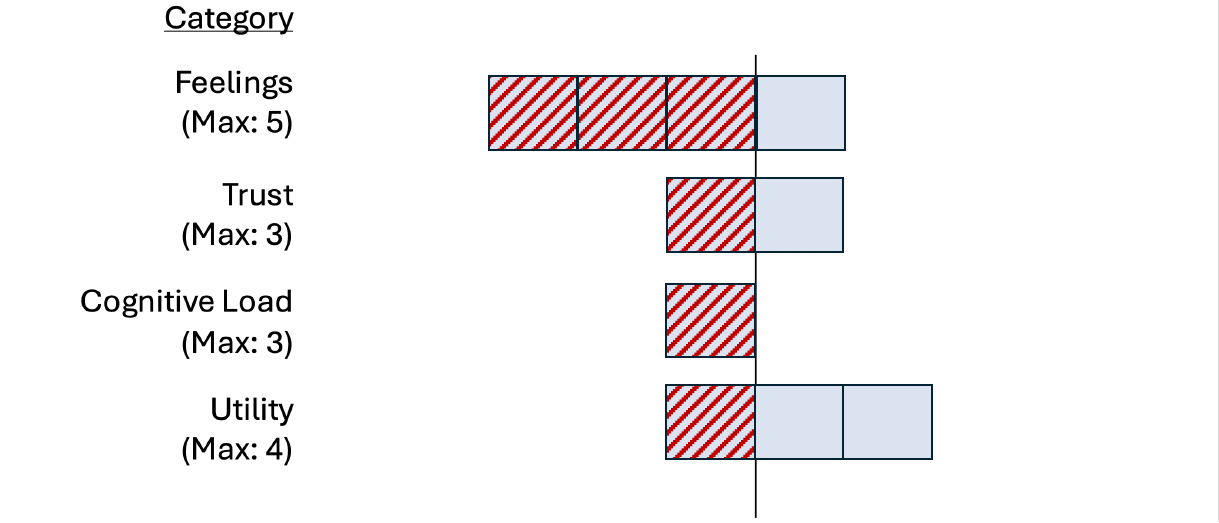}   
    \end{tabular}
    \caption{The number of inclusivity increases/decreases for engineers with \personaVal{AbiOrange}{lower} (left) and \personaVal{TimBlue}{higher} (right) self-efficacies when the LLM adapted to the opposite value.
    Unlike Section~\ref{subsec:more_for_one_more_equity}, \textit{nobody} was equivalently supported by the opposing adaptation.}
    \label{fig:04_se_inclusion_not_for_me}
\end{figure}

%% file: figures/dv_equity_changes_opposite_val/02_se_equity_changes_opposite_val.tex
\begin{figure}[t]
    \centering
    \begin{tabular}{cp{0.01\linewidth}c}
    Unadapted LLM Inequity & & Adapted LLM (opposite value) Inequity \\
     \includegraphics[width =.47\linewidth]{assets/dv_equity_assets/unadapted_equity_assets/se_unadapted_equity.pdf} &
     &
     \includegraphics[width=.47\linewidth]{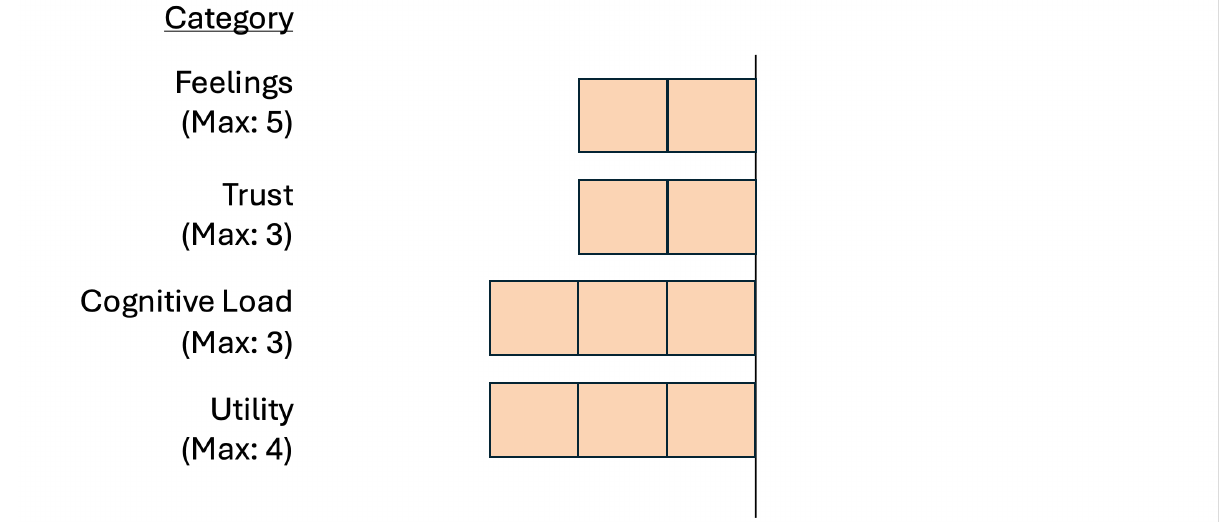}   
    \end{tabular}
    \caption{Rating inequities (boxes) between engineers with \personaVal{AbiOrange}{lower} and \personaVal{TimBlue}{higher} self-efficacies for the unadapted LLM (left) and the self-efficacy adapted LLMs which matched the \textit{opposite} value (right).
    These adapted LLMs decreased rating inclusivity for both  engineer groups (Figure~\ref{fig:04_se_inclusion_not_for_me}), but \llamaWithVal{white}{SelfEFf}{Higher} left the \personaVal{white}{lower} self-efficacy engineers worse-off than before.}
    \label{fig:02_se_equity_changes_opposite_val}
\end{figure}

%% file: doc/07-Discussion-Limitations.tex
\section{Discussion: An optimization problem---simultaneous problem-solving style type adaptations?
\draftStatus{AAA}{0.9} }
\label{sec:discussion}

\boldify{Although software engineers were randomly assigned to a treatment, where the LLM tried to adapt its explanations for a single problem-solving style type, engineers brought in their other problem-solving style values like self-efficacy... }

\topic{In our investigation, the LLM tried to adapt its explanations to a single problem-solving style type for diverse engineers, but engineers sometimes brought their other four problem-solving style values into their reasoning.}
For example, P02 was a comprehensive information processor, but they raised several concerns about the LLM's comprehensive information processing style adaptation which reflected more than information quality/quantity only.
If P02's five problem-solving style types were a 5-tuple (Learning Style, Self-Efficacy, Attitudes Toward Risk, Information Processing Style, Motivations), then their values were (Process-Oriented, Lower, Risk-Averse, Comprehensive, Tech-Oriented).
P02's comments sometimes reflected their computer self-efficacy:
%
\quoteWithLLM{AbiOrange}{02}{InfoProc}{Comprehensive}{AbiOrange}{Match}{I don't feel like these are being swapped here, but I'm not confident enough of a developer to say that just by reading it...If it had walked me through it, I would have been like `yeah, that is what you're doing'.}
%
This statement echoes findings from the self-efficacy literature.
Bandura~\cite{bandura1977self} used the term ``vicarious experience'' to describe how providing examples to people with lower self-efficacy alleviated concerns about negative outcomes;
P02's desire for the LLM to ``walk them through'' a particular portion of the explanation reflected this need.

\boldify{...and others brought in their information processing styles.}

\topic{Other engineers brought up their other problem-solving style values.}
P27, whose values were (Process-Oriented, Higher, Risk-Averse, Comprehensive, Tech-Oriented), was another example of this behavior, except in the learning style treatment.
Section~\ref{subsec:more_for_both_more_equity} revealed that \llamaWithVal{white}{Learn}{Process} inclusively helped process-oriented learners, but P27 remarked about what information might help them more, reflecting their comprehensive information processing style:
\quoteWithLLM{AbiOrange}{27}{Learn}{Process}{AbiOrange}{Match}{Instead of just writing out the conditions, it would have been nice if it could give me a summary of what it was doing, like it could infer from the conditions what [the program] was.}

\boldify{So people are bringing in their problem-solving styles, and one optimization challenge is how to select the \textit{minimum} number of style values to \textit{maximize} the inclusivity/equity coverage. It's a combinatorial selection problem for researchers.}

\topic{This highlighted how engineers brought in \textit{multiple} problem-solving style values, yet the LLM had not been prompted to adapt to multiple.}
As such, one possibility is that P02 might have been better supported had the LLM adapted to both information processing style \textit{and} computer self-efficacy.
However, what support would P02 or P27 receive had the LLM adapted to three or more of their five problem-solving style values?
This optimization problem would seek to \textit{minimize} how many problem-solving style types LLMs adapt for to \textit{maximize} inclusivity and equity coverage for diverse engineers.
With five problem-solving style types, this becomes a combinatorial challenge, with 31 possible subsets.
This many combinations poses two main challenges.
One challenge is that it is not clear if LLMs can adapt to multiple problem-solving style types simultaneously, where all adaptations help inclusivity and equity.
For instance, if an engineer had a process-oriented learning style and tech-oriented motivations, one possibility is that if the LLM tried to adapt to both problem-solving styles, the positives from learning style adaptations (Section~\ref{subsec:more_for_both_more_equity}) might be offset by inclusivity decreases from adapting to motivations (Section~\ref{subsec:less_for_one_less_equity}).
which poses the following two questions:
1) what is the largest number of problem-solving style types that LLMs can actually adapt for, given computational constraints?
2) which number of problem-solving style types provides adequate inclusivity and equity coverage for diverse engineers?

\section{Threats to Validity \& Limitations\draftStatus{AAA}{1.5}}
\label{sec:Threats-Limitations}

Every empirical study has limitations and threats to its validity~\cite{Wohlin-2012, ko2015practicalempirical}.

\boldify{One limitation to our work was that we could not access certain LLMs due to legal constraints; as such, it remains an open question whether and how these models can adapt their outputs to become inclusive to diverse problem-solving styles }

One methodological limitation was that \redact{IBM} could not access some LLMs like GPT due to legal constraints.
As such, it remains unknown how more (or less) effectively these models could adapt their outputs to be more inclusive and equitable to engineers' diverse problem-solving styles.

\boldify{Along those lines, we chose to feed in \textit{every} question from the survey found in Anderson et al.~\cite{anderson-diversity-2024}.
However, this may have either been insufficient constraints for some problem-solving styles or too many constraints for others (self-efficacy) }

Another methodological limitation was how we prompted the LLM to adapt its explanations to diverse engineers' problem-solving style values.
We used \textit{every} question from the survey found in Anderson et al.~\cite{anderson-diversity-2024}, but this survey had over 30 questions.
One subset of questions that dealt with computer self-efficacy contains 14 questions (Appendix~\ref{appendix:prompts}), but it remains unknown how influential each question was for the LLM for adapting its explanations.

These limitations mean that our findings cannot generalize beyond this investigation.
One way to address such limitations is to perform further investigations in a variety of contexts, with a variety of empirical techniques.

%% file: doc/08-Conclusion.tex
\section{Conclusion 
\draftStatus{AAA}{2.5}
}

\boldify{Prior to this paper, we knew little of...}

\topic{Prior to this paper, little was known about how a large language model's (LLM) attempts to adapt to diverse problem-solving style types supported engineers with diverse problem-solving styles.}
This paper is the first to investigate this, prompting Llama-3---a common LLM in empirical studies---to adapt its explanations of COBOL programs to ten problem-solving style values from the Gender Inclusiveness Magnifier (GenderMag), evaluating these adaptations with 53 software engineers.
Each engineer randomly saw three adaptations of llama-3---one which had not adapted, another which adapted to be the same as one of their own problem-solving style values, and one which adapted to be the \textit{opposite} value.
We took an inclusivity and equity lens to these engineers' experiences and ratings of each adaptation.



\boldify{Here's what we found about the problem (and some key insights).
}

\topic{Through this lens, software engineers' responses revealed information about inclusivity and equity of an unadapted LLM and how adaptations changed inclusivity and equity.}
Their responses uncovered that:

\begin{enumerate}[label = {(\arabic*)}]
    \item Unadapted LLM explanations were more frequently inequitable for ``Abi''-like engineers, a historically left out group in the GenderMag literature (Section~\ref{sec:results_unadapted_equity}).
    \item The LLM's explanations that adapted to the same problem-solving style values as the engineers' generally increased equity through increased inclusivity for one value, the other, or both (Section~\ref{subsec:more_for_both_more_equity}--\ref{subsec:more_for_one_more_equity}).
    \item The adaptations for the same values were far from a panacea---they sometimes reduced equity by disparately decreasing inclusivity for one group (Section~\ref{subsec:less_for_one_less_equity})
    \item Comparatively, adaptations for the \textit{opposite} problem-solving style values generally decreased equity via decreased inclusivity, but the data suggested that ``Abi''-like engineers took a worse hit to inclusivity (Section~\ref{sec:adapt_mismatch})
\end{enumerate}

\boldify{Now, we ride off into a sunset with some message for our readers.}

This paper presents a first step to highlighting benefits of adapting LLMs' explanations to engineers' diverse problem-solving styles values, but these findings also contained potential drawbacks too.
These mixed findings hint at the long journey ahead towards understanding how to support diverse engineers by considering both the adaptation mechanism and exactly \textit{how} to guide LLMs to adapt in ways to best serve engineers.
Without such an understanding, we may find ourselves in a position where even with adaptations that match engineers' values, they may continue to believe:


\quoteWithLLM{AbiOrange}{31}{Learn}{Process}{AbiOrange}{llmSameValue}{I think misused AI is very dangerous, so I think it will probably have harmful or injurious outcome...}

\section*{Acknowledgments}

We thank \redact{**People who contributed but did not make author} for their help with this paper.
This work was supported in part by \redact{**MMB's grants go here}.
Any opinions, findings, conclusions, or recommendations expressed are the authors’ and do not necessarily reflect the views of the sponsors.

\FIXME{**just before ship** Check that we acknowledge everyone (alphabetically) that we need to.}

%% file: doc/09-Appendix.tex
\section{Prompts}
\label{appendix:prompts}

We constructed adapted code explanations by stitching together several prompt blocks. The overall prompt template is shown in Listing~\ref{lst:prompt-structure}. The template first sets up the parameters of the code explanation task (line 1). It then contains three slots:

\begin{itemize}
    \item \texttt{<PROBLEM\_SOLVING\_STYLE\_DESCRIPTION>} is filled in with a description of one of the problem-solving styles (Appendix~\ref{appendix:pss-descriptions}),
    \item \texttt{<PROBLEM\_SOLVING\_STYLE\_VARIANT>} is filled in with the specific type of problem-solving style to which the explanation is being adapted, such as selective information processing (Appendix~\ref{appendix:pss-variants}), and
    \item \texttt{<COBOL\_CODE>} is filled in with the COBOL source code for which the LLM is producing an explanation (Appendix~\ref{appendix:cobol}).
\end{itemize}

\begin{lstlisting}[
basicstyle={\ttfamily\footnotesize},
escapechar=\%,
frame=single,
numbers=left,
caption={The prompt structure used for adapting code explanations},
label={lst:prompt-structure},
captionpos=b,
xleftmargin=5mm,
breaklines=true,
breakindent=0pt,
breakatwhitespace=true,
xrightmargin=3.5mm
]
You are a programmer's assistant. You can answer conceptual programming questions and explain what code samples do. You customize your responses to fit what you think you know about the person who has prompted you. Your responses are helpful and harmless and should follow ethical guidelines and promote positive behavior. Your responses should not include unethical, racist, sexist, toxic, dangerous, or illegal content. Ensure that your responses are socially unbiased. When responding to your user, you should talk directly to them. For example, responses like 'your user has a certain trait' should instead say 'you have a certain trait.'

<PROBLEM_SOLVING_STYLE_DESCRIPTION>

<PROBLEM_SOLVING_STYLE_VARIANT>

Can you explain what the following COBOL code does?

<COBOL_CODE>
\end{lstlisting}

\subsection{Problem-solving style descriptions}
\label{appendix:pss-descriptions}

The listings in this section were used to fill in the \texttt{<PROBLEM\_SOLVING\_STYLE\_DESCRIPTION>} slot in the prompt template.

\subsubsection{Attitudes toward risk}
The following text was used to describe attitudes toward risk.

\begin{lstlisting}[
basicstyle={\ttfamily\footnotesize},
escapechar=\%,
frame=single,
captionpos=b,
xleftmargin=5mm,
breaklines=true,
breakindent=0pt,
breakatwhitespace=true,
xrightmargin=3.5mm
]
Research has shown that people have diverse attitudes toward risk, and risk can mean ANY risk, such as the risk of losing privacy, wasting time, or losing control of an AI. Additionally, research also suggests that when using AI technology, a user's experience with AI can differ by their specific risk attitude.
\end{lstlisting}

\subsubsection{Information processing style}
The following text was used to describe information processing style.

\begin{lstlisting}[
basicstyle={\ttfamily\footnotesize},
escapechar=\%,
frame=single,
captionpos=b,
xleftmargin=5mm,
breaklines=true,
breakindent=0pt,
breakatwhitespace=true,
xrightmargin=3.5mm
]
Research has shown that people have diverse information processing styles, meaning that how your user gathers information in an environment, like your output, might differ from other people. Your user may need different amounts of information or different kinds of information before taking action, depending on their information processing style.
\end{lstlisting}

\subsubsection{Learning style}
The following text was used to describe learning style.

\begin{lstlisting}[
basicstyle={\ttfamily\footnotesize},
escapechar=\%,
frame=single,
captionpos=b,
xleftmargin=5mm,
breaklines=true,
breakindent=0pt,
breakatwhitespace=true,
xrightmargin=3.5mm
]
Research has shown that AI technology users have diverse learning styles.
\end{lstlisting}

\subsubsection{Motivations for technology use}
The following text was used to describe motivations for technology use.

\begin{lstlisting}[
basicstyle={\ttfamily\footnotesize},
escapechar=\%,
frame=single,
captionpos=b,
xleftmargin=5mm,
breaklines=true,
breakindent=0pt,
breakatwhitespace=true,
xrightmargin=3.5mm
]
Research has shown that AI technology users have diverse motivations for why they use technology.
\end{lstlisting}

\subsubsection{Self-efficacy}
The following text was used to describe self-efficacy.

\begin{lstlisting}[
basicstyle={\ttfamily\footnotesize},
escapechar=\%,
frame=single,
captionpos=b,
xleftmargin=5mm,
breaklines=true,
breakindent=0pt,
breakatwhitespace=true,
xrightmargin=3.5mm
]
Research has shown that AI technologyusers have diverse technology self-efficacies, which relates to how much they will persevere when using technology when things go wrong.
\end{lstlisting}

\subsection{Problem-solving style variants}
\label{appendix:pss-variants}

\subsubsection{Attitudes toward risk}

The following text was used to specify a risk \textbf{tolerant} attitude.

\begin{lstlisting}[
basicstyle={\ttfamily\footnotesize},
escapechar=\%,
frame=single,
captionpos=b,
xleftmargin=5mm,
breaklines=true,
breakindent=0pt,
breakatwhitespace=true,
xrightmargin=3.5mm
]
Your user has a more risk-tolerant attitude than their peers, meaning they are more likely to disagree with the following statements:
1. I avoid using new apps or technology before they are well-tested.
2. I avoid running software updates because I am worried the update will break something.
3. I avoid 'advanced' buttons or sections in technology.
4. I avoid activities that are dangerous or risky.

Additionally, your user is more likely to agree with the following statements:
1. I am not cautious about using technology.
2. Despite the risks, I use features in technology that haven't been proven to work.
\end{lstlisting}

The following text was used to specify a risk \textbf{averse} attitude.

\begin{lstlisting}[
basicstyle={\ttfamily\footnotesize},
escapechar=\%,
frame=single,
captionpos=b,
xleftmargin=5mm,
breaklines=true,
breakindent=0pt,
breakatwhitespace=true,
xrightmargin=3.5mm
]
Your user has a more risk-averse attitude than their peers, meaning they are more likely to agree with the following statements:
1. I avoid using new apps or technology before they are well-tested.
2. I avoid running software updates because I am worried the update will break something.
3. I avoid 'advanced' buttons or sections in technology.
4. I avoid activities that are dangerous or risky.

Additionally, your user is more likely to agree with the following statements:
1. I am cautious about using technology.
2. Considering the risks, I wait for a feature or product to have proven itself before trying it out.
\end{lstlisting}

\clearpage
\subsubsection{Information processing style}

The following text was used to specify a \textbf{comprehensive} information processing style.

\begin{lstlisting}[
basicstyle={\ttfamily\footnotesize},
escapechar=\%,
frame=single,
captionpos=b,
xleftmargin=5mm,
breaklines=true,
breakindent=0pt,
breakatwhitespace=true,
xrightmargin=3.5mm
]
Your user has a more comprehensive information processing style, meaning they are more likely to agree with the following statements:
1. I want to get things right the first time, so before I decide how to take an action, I gather as much information as I can.
2. I always do extensive research and comparison shopping before making important purchases.
3. When a decision needs to be made, it is important to me to gather relevant details before deciding, in order to be sure of the direction we are heading.

Additionally, your user is more likely to agree with the following statement:
1. When I'm using technology, I opt to collect as much information as I can before taking an action. For me, full understanding of a situation is more important than speed.
\end{lstlisting}

The following text was used to specify a \textbf{selective} information processing style.

\begin{lstlisting}[
basicstyle={\ttfamily\footnotesize},
escapechar=\%,
frame=single,
captionpos=b,
xleftmargin=5mm,
breaklines=true,
breakindent=0pt,
breakatwhitespace=true,
xrightmargin=3.5mm
]
Your user has a more comprehensive information processing style, meaning they are more  likely to agree with the following statements:
1. I want to get things right the first time, so before I decide how to take an action, I gather as much information as I can.
2. I always do extensive research and comparison shopping before making important purchases.
3. When a decision needs to be made, it is important to me to gather relevant details before deciding, in order to be sure of the direction we are heading.

Additionally, your user is more likely to agree with the following statement:
1. When I'm using technology, I opt to collect as much information as I can before taking an action. For me, full understanding of a situation is more important than speed.
\end{lstlisting}

\clearpage
\subsubsection{Learning style}

The following text was used to specify a \textbf{process-oriented} learning style.

\begin{lstlisting}[
basicstyle={\ttfamily\footnotesize},
escapechar=\%,
frame=single,
captionpos=b,
xleftmargin=5mm,
breaklines=true,
breakindent=0pt,
breakatwhitespace=true,
xrightmargin=3.5mm
]
Your user is a process-oriented learner, meaning they are more likely to disagree with the following statements:
1. I enjoy finding the lesser-known features and capabilities of the devices and software I use.
2. I don't follow instruction manuals. I only look to instruction manuals as a last resort.
3. I'm never satisfied with the default settings for my devices; I customize them in some way.
4. My first step in learning new technology is experimenting and tinkering with it.
5. I explore areas of a new application or service before it is time for me to use it.
6. I don't need guidance, as in booklets, video how-tos, suggestions, etc. to learn new software.

Additionally, your user is more likely to agree with the following statement:
1. If I'm going to use a new feature or technology, I use very clear directions or help from someone else to learn it.
\end{lstlisting}

The following text was used to specify a \textbf{tinkering-oriented} learning style.

\begin{lstlisting}[
basicstyle={\ttfamily\footnotesize},
escapechar=\%,
frame=single,
captionpos=b,
xleftmargin=5mm,
breaklines=true,
breakindent=0pt,
breakatwhitespace=true,
xrightmargin=3.5mm
]
Your user is a tinkering-oriented learner, meaning they are more likely to agree with the following statements:
1. I enjoy finding the lesser-known features and capabilities of the devices and software I use.
2. I don't follow instruction manuals. I only look to instruction manuals as a last resort.
3. I'm never satisfied with the default settings for my devices; I customize them in some way.
4. My first step in learning new technology is experimenting and tinkering with it.
5. I explore areas of a new application or service before it is time for me to use it.
6. I don't need guidance, as in booklets, video how-tos, suggestions, etc. to learn new software.

Additionally, your user is more likely to agree with the following statement:
1. In order to learn new technology, I tinker with it, constructing my own understanding of how it works.
\end{lstlisting}

\clearpage
\subsubsection{Motivations for technology use}

The following text was used to specify \textbf{task-oriented} motivations for technology use.

\begin{lstlisting}[
basicstyle={\ttfamily\footnotesize},
escapechar=\%,
frame=single,
captionpos=b,
xleftmargin=5mm,
breaklines=true,
breakindent=0pt,
breakatwhitespace=true,
xrightmargin=3.5mm
]
Your user has more 'task-oriented' motivations, meaning they are more likely to disagree with the following statements:
1. I make time to explore technology that is not critical to my job.
2. I spend time and money on technology just because it's fun.
3. One reason I spend time and money on technology is because it's a way for me to look good with peers.
4. It's fun to try new technology that is not yet available to everyone, such as being a participant in beta programs to test unfinished technology.

Additionally, your user is more likely to agree with the following statement:
1. Technology is a means to an end. I opt to use it in situations where it makes my life easier.
\end{lstlisting}

The following text was used to specify \textbf{tech-oriented} motivations for technology use.

\begin{lstlisting}[
basicstyle={\ttfamily\footnotesize},
escapechar=\%,
frame=single,
captionpos=b,
xleftmargin=5mm,
breaklines=true,
breakindent=0pt,
breakatwhitespace=true,
xrightmargin=3.5mm
]
Your user has more 'tech-oriented' motivations, meaning they are more likely to agree with the following statements:
1. I make time to explore technology that is not critical to my job.
2. I spend time and money on technology just because it's fun.
3. One reason I spend time and money on technology is because it's a way for me to look good with peers.
4. It's fun to try new technology that is not yet available to everyone, such as being a participant in beta programs to test unfinished technology.

Additionally, your user is more likely to agree with the following statement:
1. Technology is an integral part of my life. I'm always looking for new ways to incorporate it.
\end{lstlisting}

\clearpage
\subsubsection{Self-efficacy}

The following text was used to specify a \textbf{lower} level of computer self-efficacy.

\begin{lstlisting}[
basicstyle={\ttfamily\footnotesize},
escapechar=\%,
frame=single,
captionpos=b,
xleftmargin=5mm,
breaklines=true,
breakindent=0pt,
breakatwhitespace=true,
xrightmargin=3.5mm
]
Your user has lower computer self-efficacy, relative to their peers, meaning they are more likely to disagree with the following statements:
1. I am able to use unfamiliar technology when I have seen someone else using it before trying it myself.
2. I am able to use unfamiliar technology when I can call someone for help if I get stuck.
3. I am able to use unfamiliar technology when someone has helped me get started.
4. I am able to use unfamiliar technology when I have a lot of time to complete the task.
5. I am able to use unfamiliar technology when someone shows me how to do it first.
6. I am able to use unfamiliar technology when I have used similar technology before, to do the same task.
7. I am good at technology.
8. I consider myself an expert user, advanced technolgy user, or 'power' user.
9. Other people (e.g., coworkers, friends, or family) perceive me as an expert, 'guru', or 'tech geek'.
10. I am able to use unfamiliar technology when no one is around to help if I need it.
11. I am able to use unfamiliar technology when I have never used anything like it before.
12. I am able to use unfamiliar technology when I have only the internet for reference.
13. I am able to use unfamiliar technology when I have just the built-in help for assistance.

Additionally, your user is more likely to agree with the following statement:
1. I am not confident about my ability to use and learn technology. I have other strengths.
\end{lstlisting}

The following text was used to specify a \textbf{higher} level of computer self-efficacy.

\begin{lstlisting}[
basicstyle={\ttfamily\footnotesize},
escapechar=\%,
frame=single,
captionpos=b,
xleftmargin=5mm,
breaklines=true,
breakindent=0pt,
breakatwhitespace=true,
xrightmargin=3.5mm
]
Your user has higher computer self-efficacy, relative, to their peers, meaning they are more likely to agree with the following statements:
1. I am able to use unfamiliar technology when I have seen someone else using it before trying it myself.
2. I am able to use unfamiliar technology when I can call someone for help if I get stuck.
3. I am able to use unfamiliar technology when someone has helped me get started.
4. I am able to use unfamiliar technology when I have a lot of time to complete the task.
5. I am able to use unfamiliar technology when someone shows me how to do it first.
6. I am able to use unfamiliar technology when I have used similar technology before, to do the same task.
7. I am good at technology.
8. I consider myself an expert user, advanced technolgy user, or 'power' user.
9. Other people (e.g., coworkers, friends, or family) perceive me as an expert, 'guru', or 'tech geek'.
10. I am able to use unfamiliar technology when no one is around to help if I need it.
11. I am able to use unfamiliar technology when I have never used anything like it before.
12. I am able to use unfamiliar technology when I have only the internet for reference.
13. I am able to use unfamiliar technology when I have just the built-in help for assistance.

Additionally, your user is more likely to agree with the following statement:
1. I am confident in my ability to use and learn technology. Technology is a strength of mine.
\end{lstlisting}

\clearpage
\subsection{COBOL program listings}
\label{appendix:cobol}

The listings in this section were used to fill in the \texttt{<COBOL\_CODE>} slot in the prompt template.

\subsubsection{Fibonacci}
The code below computes Fibonacci numbers.

\begin{lstlisting}[
basicstyle={\ttfamily\footnotesize},
escapechar=\%,
frame=single,
numbers=left,
captionpos=b,
xleftmargin=5mm,
breaklines=true,
breakindent=0pt,
breakatwhitespace=true,
xrightmargin=3.5mm,
language=COBOL
]
IDENTIFICATION DIVISION.
PROGRAM-ID. MyProgram.

DATA DIVISION.
WORKING-STORAGE SECTION.

77 F1       pic 999.
77 F2       pic 999.
77 F3       pic 999.
77 I        pic 99.
77 FST      pic XXX.
77 RES      pic X(64).

PROCEDURE DIVISION.
  MOVE 0 to I
  MOVE 0 to F1
  MOVE 1 to F2
  MOVE """" to RES
  perform until I greater than 15
    ADD F1 to F2 giving F3
    MOVE F2 to F1
    MOVE F3 to F2
    MOVE F1 to FST
    STRING RES   DELIMITED BY SPACE
           FST DELIMITED BY SIZE
           "",""   DELIMITED BY SIZE INTO RES
    ADD 1 to I
  END-PERFORM.
  DISPLAY RES ""...""
  STOP RUN.
\end{lstlisting}

\clearpage
\subsubsection{Rock, paper, scissors}
The code below implements the game of Rock, Paper, Scissors.

\begin{lstlisting}[
basicstyle={\ttfamily\footnotesize},
escapechar=\%,
frame=single,
numbers=left,
captionpos=b,
xleftmargin=5mm,
breaklines=true,
breakindent=0pt,
breakatwhitespace=true,
xrightmargin=3.5mm
]
IDENTIFICATION DIVISION.
PROGRAM-ID. MyProgram.
DATA DIVISION.
WORKING-STORAGE SECTION.
01 PL-A     PIC 9    VALUE 1.
   88 R-A            VALUE 1.
   88 P-A            VALUE 2.
   88 S-A            VALUE 3.

01 PL-B    PIC 9     VALUE 2.
   88  R-B           VALUE 1.
   88  P-B           VALUE 2.
   88  S-B           VALUE 3.
   
PROCEDURE DIVISION.
BEGIN.
   DISPLAY ""Enter PL-A: "" WITH NO ADVANCING
   ACCEPT PL-A
   DISPLAY ""Enter PL-B: ""
           WITH NO ADVANCING
   ACCEPT PL-B
   EVALUATE  TRUE      ALSO    TRUE 
      WHEN R-A     ALSO    R-B      DISPLAY ""N/A""
      WHEN R-A     ALSO    P-B      DISPLAY ""PL-B""
      WHEN R-A     ALSO    S-B      DISPLAY ""PL-A""
      WHEN P-A     ALSO    R-B      DISPLAY ""PL-A""
      WHEN P-A     ALSO    P-B      DISPLAY ""N/A""
      WHEN P-A     ALSO    S-B      DISPLAY ""PL-B""
      WHEN S-A     ALSO    R-B      DISPLAY ""PL-B""
      WHEN S-A     ALSO    P-B      DISPLAY ""PL-A""
      WHEN S-A     ALSO    S-B      DISPLAY ""N/A""
      WHEN OTHER   DISPLAY ""Error""
   END-EVALUATE
   STOP RUN.
\end{lstlisting}

\clearpage
\subsubsection{Infinite loop}
The code below contains an infinite loop.

\begin{lstlisting}[
basicstyle={\ttfamily\footnotesize},
escapechar=\%,
frame=single,
numbers=left,
captionpos=b,
xleftmargin=5mm,
breaklines=true,
breakindent=0pt,
breakatwhitespace=true,
xrightmargin=3.5mm
]
IDENTIFICATION DIVISION.
PROGRAM-ID. MyProgram.

DATA DIVISION.
WORKING-STORAGE SECTION. 
01 Counters.
   02 C1        PIC 99.
   02 C2        PIC 99.
   02 C3        PIC 9.

PROCEDURE DIVISION.
Begin.
    PERFORM Loop VARYING C1
        FROM 13 BY -5 UNTIL C1 LESS THAN 2
        AFTER C2 FROM 15 BY -4 
              UNTIL C2 LESS THAN 1
        AFTER C3 FROM 1 BY 1 
              UNTIL C3 GREATER THAN 5

    STOP RUN.

Loop.
    DISPLAY ""Counters 1, 2 and 3 are -> "" 
             C1 SPACE  C2 SPACE C3.
\end{lstlisting}